\documentclass{emulateapj}

\usepackage{graphicx}
\usepackage{epstopdf}
\usepackage{amsmath}

\def\simlt{\mathrel{\hbox{\rlap{\hbox{\lower4pt\hbox{$\sim$}}}\hbox{$<$}}}}
\def\simgt{\mathrel{\hbox{\rlap{\hbox{\lower4pt\hbox{$\sim$}}}\hbox{$>$}}}}

\def\ale{\mathrel{\hbox{\rlap{\hbox{\lower4pt\hbox{$\sim$}}}\hbox{$<$}}}}
\def\age{\mathrel{\hbox{\rlap{\hbox{\lower4pt\hbox{$\sim$}}}\hbox{$>$}}}}

\def\gs{\mathrel{\raise0.35ex\hbox{$\scriptstyle >$}\kern-0.6em
\lower0.40ex\hbox{{$\scriptstyle \sim$}}}}
\def\ls{\mathrel{\raise0.35ex\hbox{$\scriptstyle <$}\kern-0.6em
\lower0.40ex\hbox{{$\scriptstyle \sim$}}}}

\def\spose#1{\hbox to 0pt{#1\hss}}
\def\simlt{\mathrel{\spose{\lower 3pt\hbox{$\mathchar"218$}}
     \raise 2.0pt\hbox{$\mathchar"13C$}}}
\def\simgt{\mathrel{\spose{\lower 3pt\hbox{$\mathchar"218$}}
     \raise 2.0pt\hbox{$\mathchar"13E$}}}

\slugcomment{\it To be submitted to the Astrophysical Journal}

\shorttitle{Broad H$\alpha$ Emission in the Late-time Spectra of  a H-poor Superluminous Supernova}
\shortauthors{Yan et al.}

\begin{document}

\title{Detection of Broad H$\alpha$ Emission Lines in the Late-time Spectra of A Hydrogen-poor Superluminous Supernova}
%The Discovery of Late time Interaction Between the Ejecta of Hydrogen Poor Superluminous Supernovae and Circumstellar Medium}
%\title{Discovery of a Hydrogen Shell around a H-poor Superluminous Supernova}
\author{ Lin Yan$^1$,
R. Quimby$^{2,3}$, E. Ofek$^4$, A. Gal-Yam$^4$, P. Mazzali$^{5,6}$, D. Perley$^{7,8}$, P. M. Vreeswijk$^4$, G. Leloudas$^{4,8}$, A. De Cia$^4$, F. Masci$^1$, S. B. Cenko$^{9,10}$, Y. Cao$^{11}$, S. R. Kulkarni$^{11}$, P. E. Nugent$^{12,13}$, Umaa D. Rebbapragada$^{14}$, P. R. Wo\'zniak$^{15}$, O. Yaron$^4$}
\affil{$^1$ Infrared Processing and Analysis Center, California Institute of Technology, Pasadena, CA 91125, USA. Email: lyan@ipac.caltech.edu}
\affil{$^2$ Department of Astronomy, San Diego State University, San Diego, CA 92182, USA}
\affil{$^3$ Kavli IPMU (WPI), UTIAS, The University of Tokyo, Kashiwa, Chiba, 277-8583, Japan}
\affil{$^4$ Department of Particle Physics and Astrophysics, Faculty of Physics, The Weizmann Institute of Science, Rehovot 76100, Israel} 
%\affil{$^3$ Department of Astronomy, California Institute of Technology, Pasadena, CA91125, USA}
\affil{$^5$ Astrophysics Research Institute, Liverpool John Moores University, IC2, Liverpool Science Park, 146 Brownlow Hill, Liverpool L3 5RF, UK}
\affil{$^6$ Max-Planck-Institut fŸr Astrophysik, Karl-Schwarzschild-Str. 1, D-85748 Garching, Germany}
\affil{$^7$ Department of Astronomy, California Institute of Technology, Pasadena, CA 91125, USA}
\affil{$^8$ Dark Cosmology Centre, Niels Bohr Institute, University of Copenhagen, Juliane Maries Vej 30, 2100 Copenhagen, Denmark}
\affil{$^{9}$ Astrophysics Science Division, NASA Goddard Space Flight Center, Mail Code 661, Greenbelt, MD 20771, USA}
\affil{$^{10}$ Joint Space-Science Institute, University of Maryland, College Park, MD 20742, USA}
\affil{$^{11}$ Astronomy Department, California Institute of Technology, Pasadena, CA 91125, USA}
\affil{$^{12}$ Lawrence Berkeley National Laboratory, Berkeley, California 94720, USA}
\affil{$^{13}$ Astronomy Department, University of California, Berkeley, 501 Campbell Hall, Berkeley, Ca 94720, USA}
\affil{$^{14}$ Jet Propulsion Laboratory, California Institute of Technology, Pasadena, CA 91109, USA}
\affil{$^{15}$ Space and Remote Sensing, ISR-2, MS-B244
Los Alamos National Laboratory
Los Alamos, NM 87545}

\begin{abstract}

iPTF13ehe is a hydrogen-poor superluminous supernova (SLSN) at $z$\,=\,0.3434, with a slow-evolving light curve and spectral features similar to SN2007bi. It rises in $83-148$\,days to reach a peak bolometric luminosity of $\sim$\,1.3$\times10^{44}$\,erg\,s$^{-1}$, then decays slowly at 0.015\,magnitude per day. The measured ejecta velocity is $\sim$\,13000\,km\,s$^{-1}$. The inferred explosion characteristics, such as the ejecta mass (70\,-\,220$M_\odot$), the total radiative and kinetic energy ($E_{rad}$\,$\sim$\,$10^{51}$\,erg, $E_{kin}$\,$\sim$\,2$\times10^{53}$\,erg), is typical of slow-evolving H-poor SLSN events. However, the late-time spectrum taken at +251\,days (rest, post-peak) reveals a Balmer H$\alpha$ emission feature with broad and narrow components, which has never been detected before among other H-poor SLSNe.  The broad component has a velocity width of $\sim$4500\,km\,s$^{-1}$ and a $\sim$300\,km\,s$^{-1}$ blue-ward shift relative to the narrow component.  We interpret this broad H$\alpha$ emission with luminosity of $\sim$2$\times10^{41}$\,erg\,s$^{-1}$ as resulting from the interaction between the supernova ejecta and a discrete H-rich shell, located at a distance of $\sim4\times10^{16}$\,cm from the explosion site.  This interaction causes the rest-frame $r$-band LC to brighten at late times.  The fact that the late-time spectra are not completely absorbed by the shock ionized H-shell implies that its Thomson scattering optical depth is likely $\leq1$, thus setting upper limits on the shell mass $\leq30M_\odot$. 
Of the existing models, a Pulsational Pair Instability Supernova (PPISN) model can naturally explain the observed 30$M_\odot$ H-shell, ejected from a progenitor star with an initial mass of $(95-150)M_\odot$ about 40\,years ago. 
We estimate that at least $\sim$15\%\ of all SLSNe-I may have late-time Balmer emission lines.

\end{abstract}

\keywords{Stars: supernova, massive stars}

\section{Introduction}

In the past decade, studies of superluminous supernovae \citep[SLSN;][]{Gal-yam2012} have flourished because of the significant increase in the number of discoveries from the new generations of deeper and wider transient surveys, such as the Palomar Transient Factory \citep[PTF;][]{law2009, rau2009}, the Panoramic Survey Telescope \&\ Rapid Response System \citep[Pan-STARRS;][]{kaiser2002}, and the Catalina Real-Time Transient Survey \citep[CRTS;][]{drake2009}.  These sources attracted a great deal of interests because of (1) their unusually high peak luminosities, brighter than $-20.5$\,mag (AB) and (2) their extremely broad light curves (LC) with very slow rise and decay rates ({\it e.g.} Nicholl et al. 2015). 
These SLSNe are $\sim$5\,-\,100 times more luminous than normal type Ia and core-collapse  SNe. 
Both unique features suggested new explosion physics and special properties of the progenitor stars. 
 
The known SLSNe can be classified into two broad categories according to their optical spectra \citep{Gal-yam2012}.
The first category shows hydrogen features, and is called SLSN-II.  The extremely large energy output and the detection of hydrogen imply that the progenitor star must have had a massive, extended H-rich envelope or circumstellar medium (CSM) when it exploded \citep{smith2007b,Chevalier2011}. The second category is comprised of SLSNe without any hydrogen in their spectra \citep[e.g.][]{quimby2011}.  It is thought that their progenitor stars have lost their hydrogen envelope long before the supernova went off.  Within this hydrogen-poor category, a sub-class, SLSN-R, displays LC that fades extremely slowly, and was proposed to be mostly powered by massive amounts of radioactive decay material. The archetypal SLSN-R is SN2007bi \citep{Gal-yam2009}.
Of all H-poor SLSNe, a small fraction is SLSN-R, and the majority of the events are classified as SLSNe-I. 
Because of the late-time spectral similarities to SNe Ic \citep{Pastorello2010}, in some papers this entire class is referred to as SLSN-Ic \citep[e.g.][]{Inserra2013, Nicholl2013}. 

Various scenarios have been proposed to explain the observed characteristics of these extremely energetic transient events. For hydrogen-poor SLSNe, it is speculated \citep{Gal-yam2009} that SLSNe-R are Pair-Instability Supernovae (PISN), as predicted theoretically in the late 1960s \citep{Rakavy1967,Barkat1967,Bond1984,Heger2002,Scannapieco2005}.  In this model, 
a progenitor star with $150M_\odot \leq {\rm M} \leq 260M_\odot$ first loses its H-envelope, and develops a massive oxygen core of 60\,-\,130$M_\odot$, which can reach well above  3$\times$$10^8$\,K ($\gamma$-ray photons). At such a high temperature, $\gamma$-ray photons start to produce electron-positron pairs.  This triggers a dramatic loss of radiative pressure, followed by rapid contraction, which then ignites burning of the He/O-core. This chain of events becomes a runaway thermonuclear explosion in only a few seconds. More importantly, the rapid burning and complete disruption of the core can also synthesize several $M_\odot$ of radioactive $^{56}$Ni, orders of magnitude more than typically seen in normal SNe.  It is this massive amount of radioactive material which was proposed to power the emission from SLSNe-R. A competing model is the spin down of a rapidly rotating, highly magnetic neutron star  \citep{Mazzali2006, Kasen2010, woosley2010} that can release enough energy to power the prolonged SLSN LC.  Some studies suggest all H-poor SLSNe can be explained by this model \citep{Inserra2013, Nicholl2013}. 
%However, some theoretical models and observations suggest that only about $\sim$10\%\ of neutron stars are born as magnetars from explosions of stars with initial masses $\sim$20$M_\odot$ \citep{Kasen2010,Davies2009}.   
Finally,  a third model is interaction powered \citep[e.g.][]{Gezari2009, Miller2009, Young2010, quimby2011, Chevalier2011,Sorokina2015} --- either the supernova ejecta interacting with a H-poor CSM, or a collision between two dense, H-poor shells previously expelled due to Pulsational Pair-Instability \citep[PPISN,][]{Woosley2007}, which arises in a progenitor star with a smaller initial mass of 95\,-\,150$M_\odot$. In this case, the He/O core is smaller, between 40\,-\,60$M_\odot$, which is massive enough to produce electron-positron pairs, but not massive enough to trigger a thermonuclear runaway explosion.  PPISN models predict multiple episodes of instabilities, which can expel the outer H-layer, followed by additional H-poor CSM shells. After enough mass is lost, the star undergoes a Fe-core collapse supernova explosion.  

For SLSN-II, the high luminosities and the slow rise/decay rates are thought to be explained by some of these four different power sources. A popular model is the interaction model, either by collisions between two dense shells -- one with H and another without --  ejected by PPISN \citep{Woosley2007}, or by strong interactions between ejecta and very dense H-rich CSM \citep{smith2007,ofek2007,Chevalier2011,Moriya2014}.   
It is important to note that a combination of these power sources --- magnetar, radioactive decay (PISN), CSM interaction and PPISN --- could work together to explain some SLSNe. 
%For example, SN2006gy is a SLSN-II, which may be powered by both CSM interaction and radioactive decay from a PISN  \citep{Smith2008}.  Alternatively, a PPISN model was also proposed to explain this same observation \citep{woosley2010}. 

The explosion physics and power sources for SLSNe could be diverse, and the associated progenitor masses could also vary from $\sim$20$M_\odot$ (magnetar, Davies et al. 2009) up to 250$M_\odot$ (PISN). 
However, what is clear is that the progenitor stars of hydrogen-poor SLSNe  must have lost most or all of their hydrogen envelopes prior to the supernova explosion. The possible ways of mass losses include massive wind and pulsational pair instabilities. 
A PPISN suggests the ejection of 10\,-\,20$M_\odot$ of H-rich material during each instability episode. This model naturally predicts that some SLSNe-I could have distant H-rich shells, previously lost due to the violent pulsational pair instabilities. At late times, the supernova ejecta would eventually run into this distant H-rich shell, and produce broad Balmer emission lines from the interaction.

In this paper, we report for the first time the observations of two hydrogen-poor SLSNe with late-time spectral signatures of the ejecta interacting with H-rich medium.  We present a detailed analysis for iPTF13ehe which has extended photometric and spectroscopic data over 400 days.  We summarize the results for the second source, PTF10aagc, at the end. 
The paper is organized as follows. The observational data is presented in \S\,\ref{sec:data}, the analysis and results are described in \S\,\ref{results}. In \S\,\ref{discuss}, we discuss the implications of these observations for various SLSN models.

Throughout the paper, we adopt a $\Lambda$CDM cosmological model with
$\Omega_{\rm{M}}$\,=\,0.286, $\Omega_{\Lambda}$\,=\,0.714, and $H_0$\,=\,69.6\,$\rm{km}\rm{s}^{-1}\rm{Mpc}^{-1}$ \citep{Planck2015}.

\section{Observations }
\label{sec:data}

iPTF13ehe was first detected as a transient source on November 25, 2013 by the intermediate Palomar Transient Factory (iPTF). Its Equatorial coordinates are RA$=$06:53:21.50 DEC$=$+67:07:56.0 (J2000).  
Using the observations presented below, we show that this event is at a redshift of 0.3434 and its photometric and spectroscopy properties are consistent with a hydrogen-poor, super luminous supernova, similar to SN2007bi. This section discusses the characteristics of the photometric and spectroscopic observations.

\subsection{Photometric data \label{data-phot}}

The iPTF13ehe photometry was obtained mostly with the PTF survey Camera \citep{Rahmer2008} on the 48\,inch Oschin Schmidt telescope (P48) and the imaging camera on the robotic 60\,inch (P60) telescope at Palomar Observatory \citep{Cenko2006}. 
Additional late-time photometry was obtained with the Large Format Camera (LFC) on the Palomar 200\,inch (P200), the Keck and the Discovery Channel Telescope (DCT).  P48 data from February 2013 set useful constraints on the explosion date. The  P48 images are processed by the PTF imaging processing pipeline written at the Infrared Processing and Analysis Center (IPAC) \citep{russ2014}. The photometry is measured using the PTF Image Differencing Extraction (PTFIDE) software \citep{frank2014}.  This package produces both Point Spread Function (PSF) fitted photometry as well as aperture photometry on the reference subtracted images. More importantly, co-added photometry and upper limits based on multi-epoch observations can be derived when the transient object is faint.

The photometry from the P60, the Keck Low Resolution Imager and Spectrograph \citep[LRIS][]{oke1995}, and the DCT are measured through an appropriate aperture with diameter (2\,-\,2.5)$\times$FWHM of the seeing disk. All photometry are in AB magnitudes, and calibrated onto the SDSS $g$, $r$ and $i$ filters.  On February 17, 2015, iPTF13ehe was observed by the HST/ACS/WFC camera in the F625W filter (PID: 13858) (De Cia et al., in prep). The supernova iPTF13ehe is clearly offset from a faint dwarf galaxy.  After the subtraction of the supernova light,  the host galaxy photometry is $24.24\pm0.06$\,magnitude (AB, $r$).  The $g$-band decline is leveling out by March 23, 2015, with a total magnitude of $24.77$.  It  faded only 0.1\,magnitude during the two months between January 22 and March, 2015.  Thus, we approximate the host brightness in $g$-band with $24.9$.   Some of the late-time  photometry is taken with LRIS Cousin $R_c$ filter.  We transformed $R_c$ magnitude to SDSS $r$ AB magnitude using the late-time spectra. The $g$,$r$ and $i$-band photometry are listed in Table 1. The tabulated photometry are not extinction corrected. 

The Schlafly \&\ Finkbeiner (2011) extinction map gives a Galactic extinction E(B$-$V)$=$0.04 at the position of iPTF13ehe. We assume the extinction law of \citep{cardelli89} with $\rm A_V/E(B - V) = 3.1$.  Specifically, ${\rm A}_g$, ${\rm A}_r$ and ${\rm A}_i$ are small, 0.14, 0.1 and 0.07\,magnitude respectively. In this paper, we have ignored any potential dust extinction from the host galaxy. This assumption is probably not too far off since most dwarf galaxies have low dust extinction and our late-time spectra do not show any significant reddening or Na\,D absorption lines. Studies of SLSN-I host galaxies also support this assumption \citep{Leloudas2015, Lunnan2014}. 

%To make $r$-band LC comparisons between different SLSNe, we need to bring all photometry to the rest-frame.  At $z$\,=\,0.3434, K-corrections are small, having very little impact on our analyses. For most supernovae, the availability of optical spectra is limited in comparison with the photometric LC.  For iPTF13ehe, we adopt the following approximation. Without actual spectra, we use $(g - r)$ colors to set the spectral slopes which determines the $r$-band K-corrections.  For our target, the derived K-corrections span a range of $-0.06$ to $-0.12$\,magnitude for $(g - r) \sim 0.3 - 0.9$. Here K-correction (KC) is defined as ${\rm M}_r = m_r - {\rm DM - KC}$, with $\rm DM$ as distance module. 

\subsection{Spectroscopy \label{data-spec}}
Spectroscopic observations of iPTF13ehe were obtained on six epochs, three near the light curve peak, and three during the late-time nebular phase ($\sim$\,300\,days since the peak).  The spectra were taken with the Keck DEep Imaging and Multi-Object Spectrograph (DEIMOS; Faber et al. 2003) and LRIS \citep{oke1995} mounted on the Keck 10\,meter telescopes, and with the Double Spectrograph (DBSP; Oke \&\ Gunn 1983) on the P200.  The spectral coverage is $\sim$\,$3000-10000$\,\AA\ and $4000-10000$\,\AA\  for LRIS+DBSP and DEIMOS respectively. The spectral resolution is moderate, with $\lambda/\delta \lambda \sim$\,$800-2000$. DEIMOS observations were reduced using the software developed by the Deep Extragalactic Evolutionary Probe 2 (DEEP2) project \citep{newman2013}.  LRIS and DBSP observations were reduced by us (D. Perley and R. Quimby) using custom written software.  The observation information is listed in Table 2.

All six spectra are flux calibrated and corrected for Galactic extinction assuming $\rm E(B-V) = 0.04$ (see \S~\ref{data-phot}). We cross-check the spectral calibration against broad band photometry. Overall, the corrections to the spectral calibrations are small.
The spectra taken near peak luminosity were not corrected for host galaxy contamination, since it is faint and negligible.  For the late-time spectra, we perform host galaxy subtraction, as described below (\S~\ref{res-spec}). All calibrated spectra will be made publicly available via WISeREP (http:/wiserep.weizmann.ac.il) \citep{Ofer2012}.

\section{Analysis and Results }
\label{results}

\subsection{Light curves: rise time  and peak bolometric luminosity \label{res-lc}}

Supernova light curves provide several important measurements which can constrain the explosion physics.  This includes three observables: the rest-frame rise time from the date of explosion to the maximum brightness ($t_{rise}^{rest}$), the peak bolometric luminosity ($L_{bol}^{peak}$) and the post-peak decay rate ($\Delta M/\Delta t$).  
%The rise time scale $t_{rise}^{rest}$ is determined as following. 

\begin{figure}[!ht]
%\plotone{/Users/lyan/Work/SLSN/13ehe/phot/obsLCv2.pdf}
\plotone{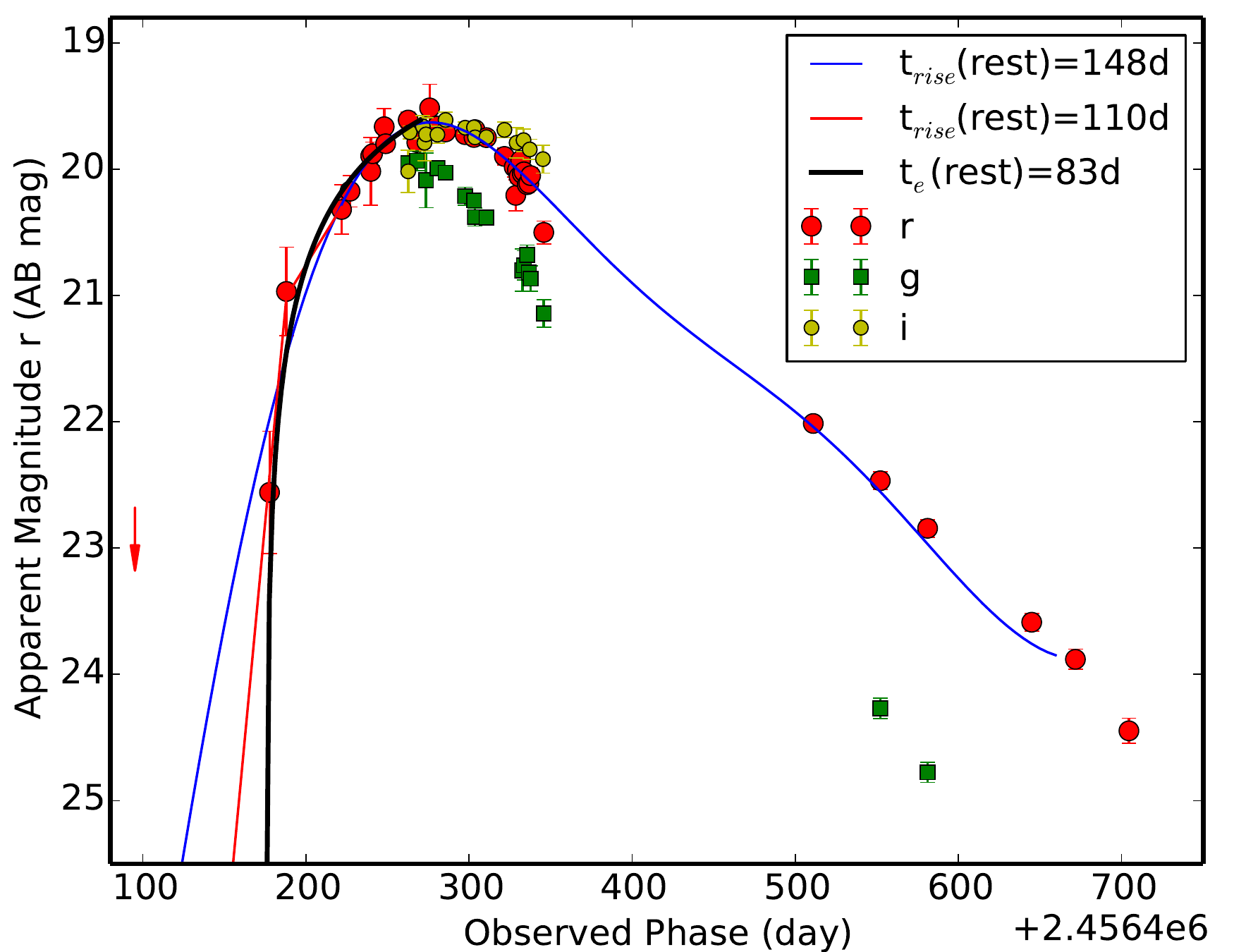}
\caption{The iPTF13ehe apparent brightness in the $g$, $r$ and $i$-band versus observed Julian date, overlaid with polynomial fits to the $r$-band LC in order to infer the peak and explosion dates.  The blue line uses all of the data, the red line uses a piece-wise polynomial fit and the black line is based on an exponential form. The time scales are in the rest-frame. \label{obslc}}
\end{figure}

Figure~\ref{obslc} illustrates the observed $g$, $r$ and $i$-band LCs as a function of Julian Date (JD).  iPTF13ehe has a total of 48 images taken eight months prior to the discovery date on November 25, 2013.  The upper limit and the two earliest detections in Figure~\ref{obslc} were derived from stacking, each using $\sim$10 images spanning over 10 days.  These early data -- often missed for SLSNe -- are very useful to constrain the rise time scale, the explosion date and for searching for SN precursors \citep[e.g.][]{Ofek2014}. The $g$ and $r$-band LCs are host-light subtracted, and the $i$-band LC covers only the peak epochs, which are not significantly affected by the faint host. 

%show a hint of a small bump in the LC before rising to its peak. This bump could be similar to what have been seen in the early observations of PTF12dam and SN2006oz \citep{Vreeswijk2015, Leloudas2012}, but much weaker than what was observed in LSQ14bdq \citep{Nicholl2015}.  These pre-peak bumps in LC could indicate interesting physics, such as recombination of a certain ionized element ({\it e.g.} He) or shock heating/cooling of CSM. However, iPTF13ehe has too few early observations for any detailed analysis.

%The $g$ and $r$-band photometry in Figure~\ref{obslc}  are host light subtracted. The $i$-band photometry covers only the peak, and is not significantly affected by the faint host.  iPTF13ehe was observed with HST/ACS/WFC camera in F625W filter (the same as SDSS-$r$) on February 17, 2015 \citep{annalisa2015}. The supernova iPTF13ehe is at the edge of the dwarf galaxy, with a separation of 0.5$^{''}$.  After the subtraction of the PSF fit to the supernova,  the host galaxy $r$-band photometry is measured as 24.24$\pm0.06$\,magnitude (AB).  We can set a limit on the $g$-band host brightness by using the color information. {\it WAITING TO SEE MAR 23,2015 $g$-photometry.}

We first determined the peak and explosion dates using polynomial fits to the $r$-band LC. iPTF13ehe reached its peak at Julian date (JD) of 2456670.77\,days.  Here we show how the explosion date, thus rise time, can be affected by various factors. 
We define the explosion date as the time when the extrapolated $r$-band magnitude is fainter than 30\,magnitude. When we use all of the data for a single polynomial fit (Figure~\ref{obslc}; blue solid line), the derived explosion date is 2456471.3\,days. But if we consider that the two earliest data points may favor a different rising slope, and fit the data with a piece-wise polynomial (red solid line), the derived explosion date is later, at JD\,=\,2456522.5\,days.  The corresponding rest-frame rise times are $t_{rise}^{rest}$\,=\,148.5 and 110\,days respectively. The large uncertainty is due to lack of early observations, and more importantly, lack of knowledge of the shape of early SLSN light curves.  For example, instead of slower rising slopes indicated by  the polynomial fit (blue and red lines in Figure~\ref{obslc}), the early LC may have a faster rising exponential form of $L(t) = L_{peak} (1.0 - e^{t_0-t \over t_e})$, as applied to a sample of SN IIn in \citet{Ofek2014}. The fit to the data yielded the explosion date $t_0$\,=\,2456575.6\,days and the exponential rise time $t_e$\,=\,83\,days (shown as black line in Figure~\ref{obslc}).  The large uncertainty in $t_{rise}^{rest}$ illustrates one critical and the most difficult aspect of supernova observations --- catch them early enough that the physical information can be derived.
We conclude that the rise time $t_{rise}^{rest}$ is in a range of $83 - 148$\,days. 

\begin{figure}[!ht]
%\plotone{/Users/lyan/Work/SLSN/13ehe/phot/absgLCv2.pdf}
\plotone{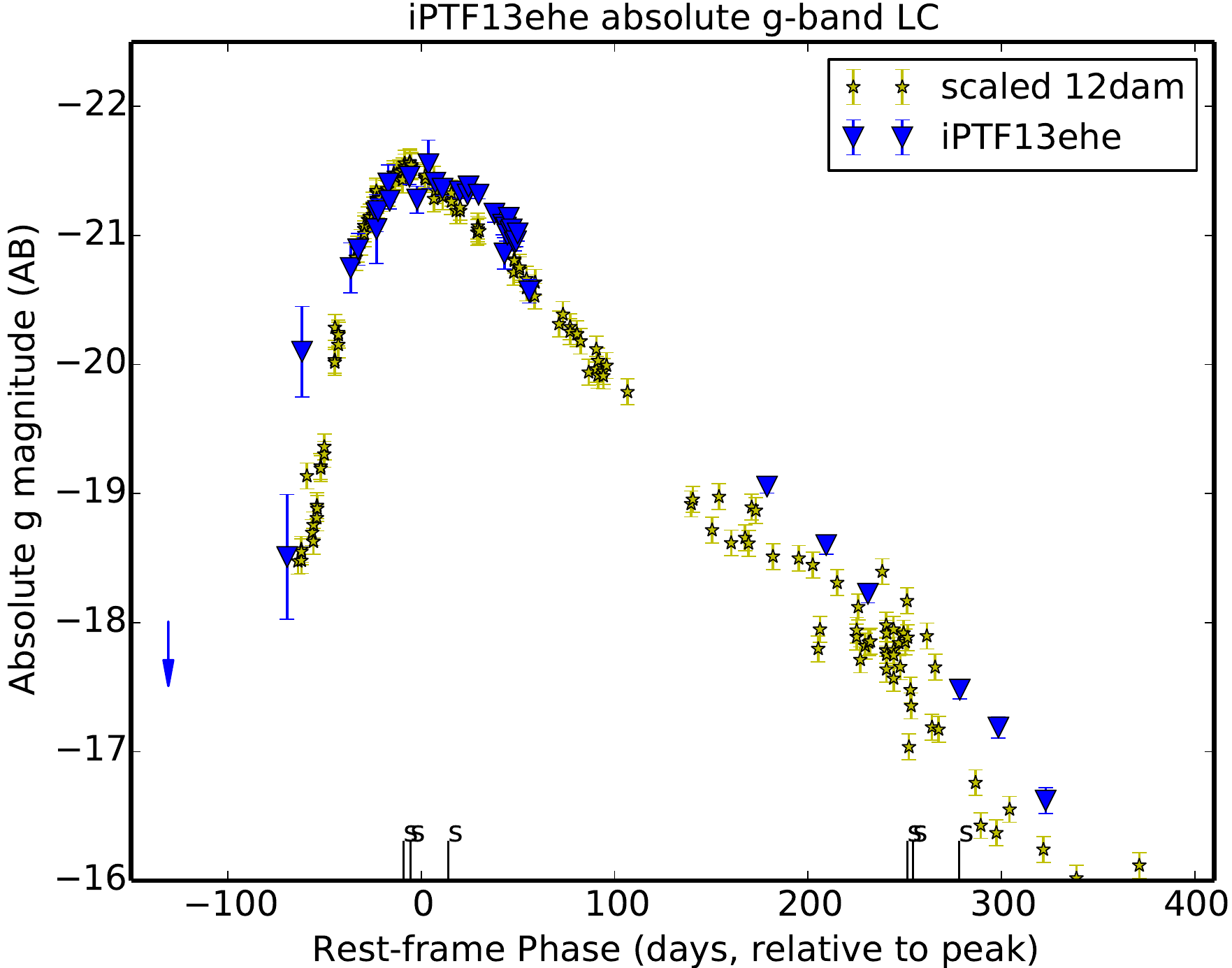}
\caption{The Comparison of the iPTF13ehe $M_g$ LC with that of PTF12dam \citep{Vreeswijk2015}. The PTF12dam $g$-band LC is shifted down by $+$0.15magnitude. \label{absglc}}
\end{figure}

Rest-frame light curves require appropriate k-corrections.  Here the k-correction, $K_{QR}$, is defined as $M_Q (rest) = m_R(obs) - DM - K_{QR}$, with the observed filter being $R$, the rest-frame filter being $Q$, and $DM$ being the distance module. We transformed the observed $r$-band LC to rest-frame $M_r$ and $M_g$ LCs by applying $K_{rr}$ and $K_{gr}$, calculated using the observed spectra at the three separated epochs (01/06/2014, 02/01/2014 and 12/17/2014).
$K_{gr}$ is $-0.28$, constant for both the early and late epochs.  The $K_{gr}$ correction is almost constant because at $z$\,=\,0.3434, the observed $r$ filter samples rest-frame 4659.8\,\AA, very close to the $g$-band $\lambda_{eff}$ at $z$\,=\,0.  Thus $K_{gr}$ is approximately $2.5 \log_{10}(1+z)$ (for details see \citealt{Hogg2002}). The derived rest-frame $M_g$ LC is shown in Figure~\ref{absglc}.

\begin{figure}[!ht]
%\plotone{/Users/lyan/Work/SLSN/13ehe/phot/absLCv4.pdf}
\plotone{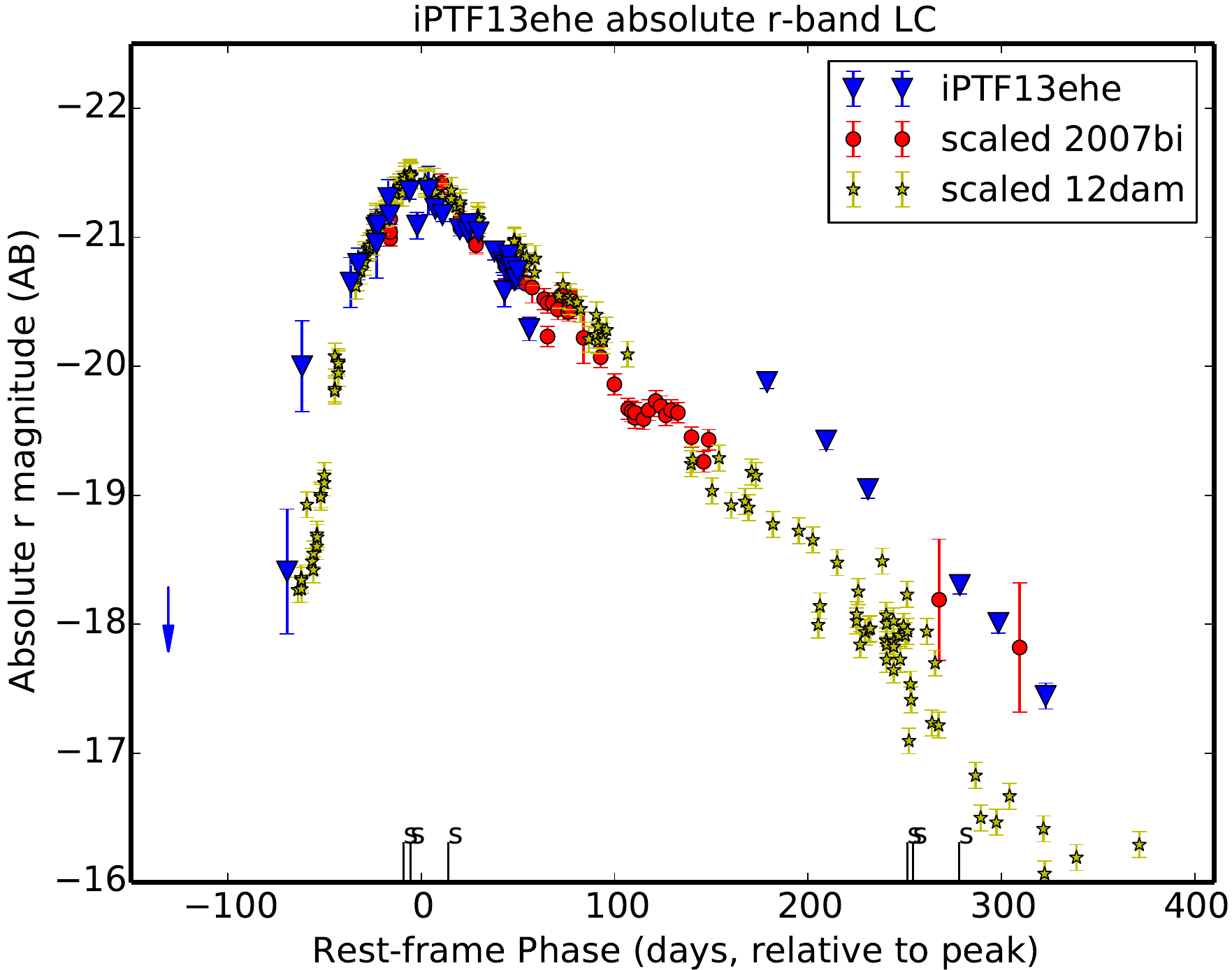}
\caption{The iPTF13ehe rest-frame, k-corrected $M_r$ light curve in comparison with those of SN2007bi and PTF12dam \citep{Gal-yam2009, Vreeswijk2015}. The scaling factor is $+$0.1\,magnitude for SN2007bi and 0.05 for PTF12dam. \label{abslc}}
\end{figure}

$K_{rr}$ is approximately $-0.38$ for the pre-peak photometry, $-0.5$ between the peak and +100\,days post-peak (rest-frame), and $+0.54$ for the rest of the late-time photometry.  We note that $K_{rr}$\,=\,+0.54 is the averaged value based on the three late-time spectra (see Figure~\ref{obslc}).  This K-correction is large and positive, making the rest-frame $M_r$ light curve brighter, as shown in Figure~\ref{abslc}.   
The change of $K_{rr}$ with time is due to (1) the emergence of a strong, broad H$\alpha$ emission line at late times (first detected at December 21, 2014); (2) the redder continuum in the supernova spectra.  The large variation in $K_{rr}$ is also responsible for the observed steep $(g-r)$ color evolution with time, from 0.3 near the peak brightness to 1.5 at later times (Figure~\ref{obslc}).  How much of the $M_r$ brightening is due to the continuum versus H$\alpha$? We test this by masking out the H$\alpha$ line from the late-time spectra and find that the calculated $K_{rr}$ from the pure continua is $+0.3$, still quite significant.

The light curves shown in Figure~\ref{abslc} and Figure~\ref{absglc} indicate a linear decline, except the brightening in the late-time $M_r$ LC.  
The decay rates are 0.0155 and 0.0149\,magnitude/day for the $r$ and $g$-band LCs respectively.
For the $r$-band LC, we measured the decay rate separately for the late-time and the post-peak photospheric period, and the values are very similar. These decline rates are much slower than that of any normal supernova in the post-peak photospheric phases.  We note it is close to the pure$^{56}$Co decay rate of 0.0097\,magnitude/day.

Figures~\ref{abslc} and \ref{absglc} compare the iPFT13ehe $M_r$ and $M_g$ LC with those of SN2007bi and PTF12dam, two well studied hydrogen-poor SLSNe \citep{Gal-yam2009, Vreeswijk2015, Nicholl2013, Chen2014}.  It is interesting to note that the decline rates of the three SLSNe are very similar, except for the elevated $M_r$ bump in iPTF13ehe after the emergence of the broad H$\alpha$ emission.  In addition, iPTF13ehe has a slower rising rate, thus a wider LC, compared to PTF12dam.  One argument against PTF12dam being a Pair Instability Supernova (PISN)  is that its LC rose much faster than model predictions \citep{Inserra2013, Nicholl2013}.

%\begin{figure}[!ht]
%\plotone{/Users/lyan/Work/SLSN/13ehe/phot/obsLC.pdf}
%\caption{The figure presents the iPTF13ehe $r$-band observed light curve. \label{obslc}}
%\end{figure}

%\begin{figure}[!ht]
%\plotone{/Users/lyan/Work/SLSN/13ehe/phot/grcolor.pdf}
%\plotone{fig4.pdf}
%\caption{The observed $(g-r)$ color evolution as a function of rest-frame time relative to the peak date.  \label{grcolor}}
%\end{figure}

\begin{figure}[!ht]
%\plotone{/Users/lyan/Work/SLSN/13ehe/spec/compspec_v3.pdf}
\plotone{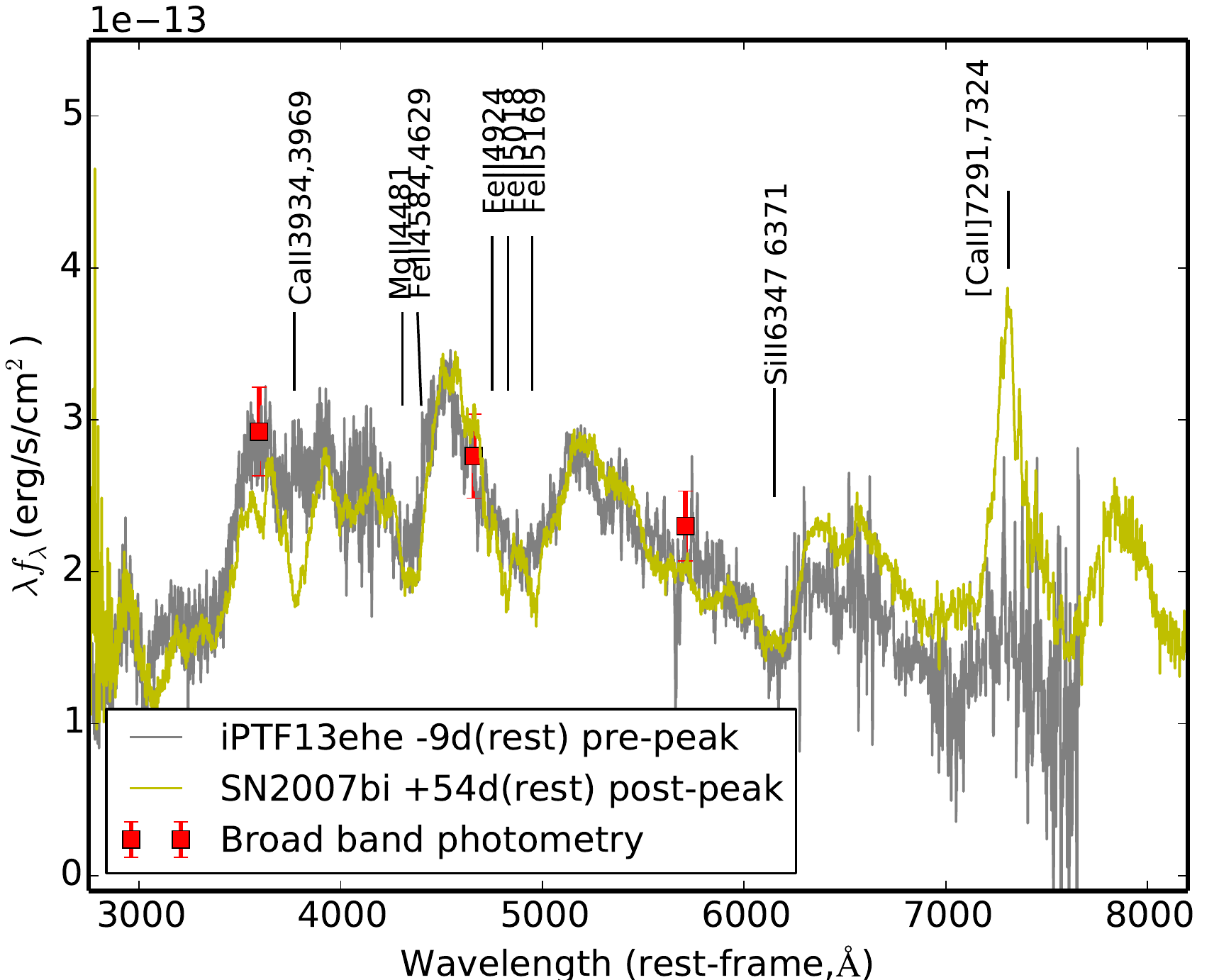}
\caption{ The optical spectrum during the photospheric phase of iPTF13ehe, taken at the pre-peak $-9$\,days.  We compare this spectrum  with that of SN2007bi, which was proposed to be a PISN and powered by a large mass of $^{56}$Ni  \citep{Gal-yam2009}. \label{obs-spec}}
\end{figure}

The peak bolometric luminosity, $L_{bol}^{peak}$, is estimated as follows.  We have $g$, $r$ and $i$ photometry taken around the time when iPTF13ehe reached its peak brightness. Thus, we have fairly good estimates of the peak fluxes in these filters. The integral over the broad band defined SED between $g$ and $i$-band results in $6\times10^{43}$\,erg/s, which sets a lower limit on $L_{bol}^{peak}$.  Figure~\ref{obs-spec} shows the spectrum in $\lambda f_\lambda$ versus $\lambda_{rest}$. This spectrum is taken at $-9$\,days pre-peak. We used the broad-band photometry taken nearest to this spectrum, and refined the flux calibration to account for slit losses.  
The continuum shape should reflect the blackbody radiation, and the bolometric flux $f_{bol}$ is simply 
\begin{equation}
\begin{split}
f_{bol}
&= \int_0^\infty ({2 \over \lambda^2}  {h\nu \over \mathrm{e}^{h\nu/k_BT} - 1})d\nu \\
&= {\sigma T^4 \over \pi} = 1.386\nu_m f(\nu_m)
\end{split}
\end{equation}

where $\nu_m$ is the frequency when $\nu f_\nu$ is at the maximum.  From the spectra shown in Figure~\ref{obs-spec},  $\lambda f_\lambda$ peaks at $\lambda$$ \sim$\,4800\AA, with $f_{bol}$\,=\,1.386*$\lambda_m f(\lambda_m)$\,=\,1.386*4800.0*4.95*$10^{-17}$\,erg/s/cm$^2$.  This yields $\rm L_{bol}^{peak}$\,=\,1.3$\times$$10^{44}$\,erg/s.  This is consistent with the lower limit set by the broad band photometry.

With $t_{rise}^{rest}$ and $L_{bol}^{peak}$, it is clear that iPTF13ehe is an extremely luminous supernova. The total radiative energy can be estimated by, $E_{rad}$\,$\geq$\,$L_{bol}^{peak}$$\times$$t_{rise}^{rest}$\,$\sim$1.3$\times10^{44}$\,erg/s\,$\times$83\,days\,$\sim9.3$$\times$$10^{50}$\,erg.

\begin{figure*}[!htb]
%\plotone{/Users/lyan/Work/SLSN/13ehe/spec/allspec.pdf}
\plotone{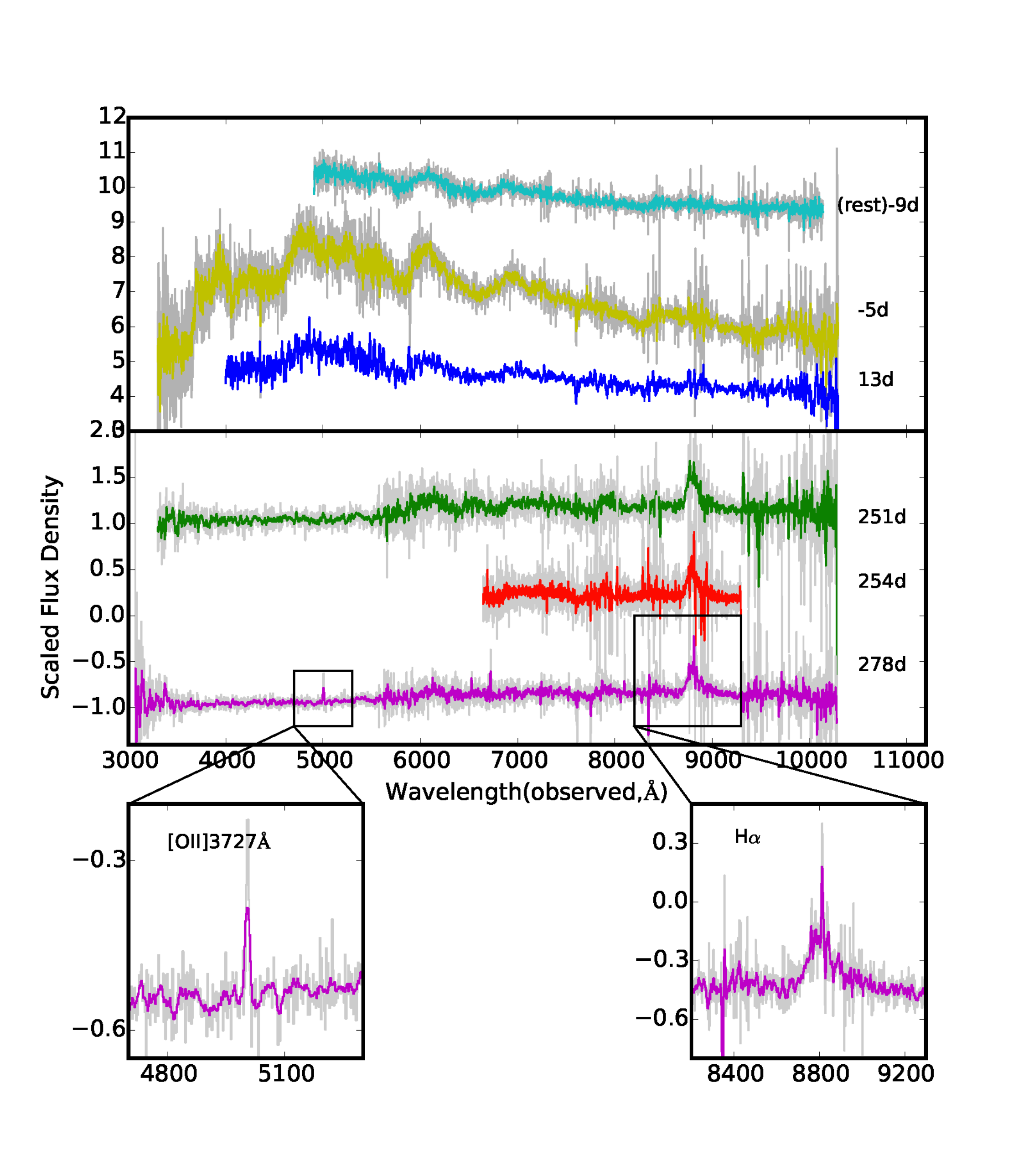}
\caption{ All six observed spectra are shown here with the rest-frame days relative to the peak brightness. For the third nebular spectrum taken at +278\,days post-peak, we detected both broad and narrow H$\alpha$ components as well as narrow [O\,II]\,3727\AA. The color lines are the smoothed spectra. See the text for details on the smoothing lengths. \label{allspec}}
\end{figure*}

\subsection{Photospheric and nebular phase spectroscopy \label{res-spec}}

iPTF13ehe was observed on six different epochs. Figure~\ref{allspec} presents these data, illustrating the evolution of the spectral features over the rest-frame time interval of 287\,days.  The color lines in the figure show the smoothed spectra, which are performed using astropy package convolve.Box1DKernel software. The smoothing length ranges from $6-8$\AA\ ($5-7$\,pixels) for the P200 and Keck LRIS spectra, and is much smaller, only $3$\AA\ (11\,pixels) for the DEIMOS high resolution spectrum (2014 Dec. 21).
In the section below, we discuss in detail the spectral properties in the photospheric and nebular phase separately.  An accurate redshift is measured using the multiple narrow emission lines detected in the final spectrum (Jan 22, 2015).

\subsubsection{Photospheric phase spectra --- Similarity between iPTF13ehe and SN2007bi \label{photo-spec}}

The earliest spectrum of iPTF13ehe, taken at $t_{rest}^{peak}$\,=$-9$ days pre-peak, is very similar to the earliest available photospheric spectrum (rest-frame +54 days post-peak) of SN2007bi, shown in Figure~\ref{obs-spec}.  No detectable H and He features are present, and the most prominent absorption features are from MgII, FeII and SiII.  This suggests that prior to the explosion, the progenitor star must have lost all of its hydrogen envelope, and the supernova explosion comes from the core containing heavier elements.  Figure~\ref{obs-spec} illustrates that the width of the blended FeII\,5169\AA\ feature is very similar for these two spectra, allowing us to roughly infer the iPTF13ehe ejecta velocity of $\sim$\,12000\,km/s.  This velocity is confirmed by other methods. For example, it is thought that the blue minimum of the  P-Cygni profile of FeII\,5169\AA\  is a fairly good indicator of the photospheric expansion velocity (Branch 2004). At observed wavelength of 6620\AA, this feature implies $v_{ej}$\,=\,$14000\pm3000$\,km/s, with the error measured from the line profile fitting. Throughout the paper, we adopt $v_{ej}$\,=\,13000\,km/s, the averaged value between 12000 and 14000\,km/s. 

The significant spectral difference between SN2007bi and iPTF13ehe is around 7322\,\AA\ where [O\,II]\,7322\AA, [Ca\,II]\,7291,7324\AA, [Fe\,II]\,7155,7172,7388,7452\AA\ emission lines are located (Figure~\ref{obs-spec}). iPTF13ehe probably has very weak Ca\,II\,3934,3969\AA\ (H \&\ K; absorption) and [Ca\,II]\,7291,7324\AA\ (emission), in contrast to SN2007bi.  Because [Ca\,II] line is sensitive to gas density and is stronger at lower density, this may indicate that the ejecta of SN2007bi contains low density regions.

A simple blackbody fitting to the spectral continuum obtained at the epoch of $-5$\,days produces a temperature $T_{BB}$ of $\sim$7000\,K. It is clear that near the peak, the optical spectra of iPTF13ehe are much flatter, {\it i.e.} cooler than those of PTF09cnd, an archetypical SLSN-I \citep{quimby2011}.  The earliest spectrum of PTF09cnd taken at $-20$\,days pre-peak has a continuum blackbody temperature of $\sim$15,000K \citep{quimby2011}. The cooler temperature in iPTF13ehe is also supported by the absence of O\,II\,4072,4415,4590\AA\ absorption features (the O$^+$ ionization potential is 35.1\,eV, 40,000\,K assuming thermal equilibrium).  In contrast, O\,II absorptions are prominent in PTF09cnd \citep{quimby2011}.  Since SN2007bi does not have pre-peak spectra, it is possible that its earlier spectra may have a hotter continuum like PTF09cnd. However, this is not the case for iPTF13ehe since the first spectra were  obtained prior to the peak epoch.  Thus, it could be that sources, like iPTF13ehe and SN2007bi, may represent a different class of hydrogen-poor super-luminous SNe, in the sense that they may have a different spectral evolution, suggesting different explosion physics.  The detailed study of this issue will be presented in Quimby et al. (in preparation), based on the full H-poor sample discovered by PTF from 2009 to 2012.  

%\begin{figure}[!ht]
%\plotone{/Users/lyan/Work/SLSN/13ehe/spec/nebular1.pdf}
%\plotone{fig7.pdf}
%\caption{The third nebular spectrum of iPTF13ehe observed at +278 days (rest-frame) post-peak.  It contains light from both the host galaxy and the supernova,  with narrow emission lines from the host galaxy and broad features from the supernova.   \label{neb1}}
%\end{figure}

\subsubsection{Nebular phase spectra --- Detection of a broad H$\alpha$ emission line \label{neb-spec}}

As Table 2 shows, we have three nebular-phase spectra taken on 
$t_{rest}^{peak}$=+251, +254 and +278 days (post-peak) or $t_{rest}^{exp}$=+322, +325 and +349\,days (from explosion JD of 2456575.6\,days) respectively.  All three spectra display a strong, broad H$\alpha$ emission line with a velocity width $>$4000\,km/s as well as a narrow H$\alpha$ emission component.  Below we discuss in turn the nature of these two components.

The first question is whether the narrow H$\alpha$ component is from the recombination of ionized hydrogen atoms in a slow moving shell or from the host galaxy.  This type of narrow emission feature is commonly seen in the spectra of SN IIn, and usually comes from the slow moving, outer layer of the H-rich circumstellar medium (CSM) surrounding the supernova. In the case of iPTF13ehe,  the complication comes from the fact that all available spectra contain signals from both the supernova and the host galaxy.  In the three nebular spectra, the narrow H$\alpha$ line is unresolved for the two spectra taken with LRIS and resolved in the high resolution DEIMOS data (FWHM of 1.8\AA) taken on December 21, 2014.  The integrated line fluxes for the narrow component are 2.2$\times10^{-17}$, 2.1$\times10^{-17}$ and 3.88$\times10^{-17}$\,erg/s/cm$^2$ for December 17, 2014, December 21, 2014 and January 22, 2015 respectively. 
The Janurary 2015 spectrum has the highest SNR and also detects H$\beta$ and narrow [O\,II]\,3727\AA\ emission lines.  The redshifts inferred independently from
the narrow H$\alpha$ and [O\,II]  lines are the same, at $z=1.3429$.  The integrated [O\,II]\,3727\AA\ line flux is 2.64$\times10^{-17}$\,erg/s/cm$^2$, implying the SFR of 0.15$M_\odot$/yr based on the conversion from Kennicutt (1998).  If all of the narrow H$\alpha$ line flux (3.88$\times10^{-17}$\,ergs/cm$^2$) comes from star formation, the inferred SFR is 0.13$M_\odot$/yr, 15\%\ lower than that inferred from [O\,II].

\begin{figure*}[!ht]
\plotone{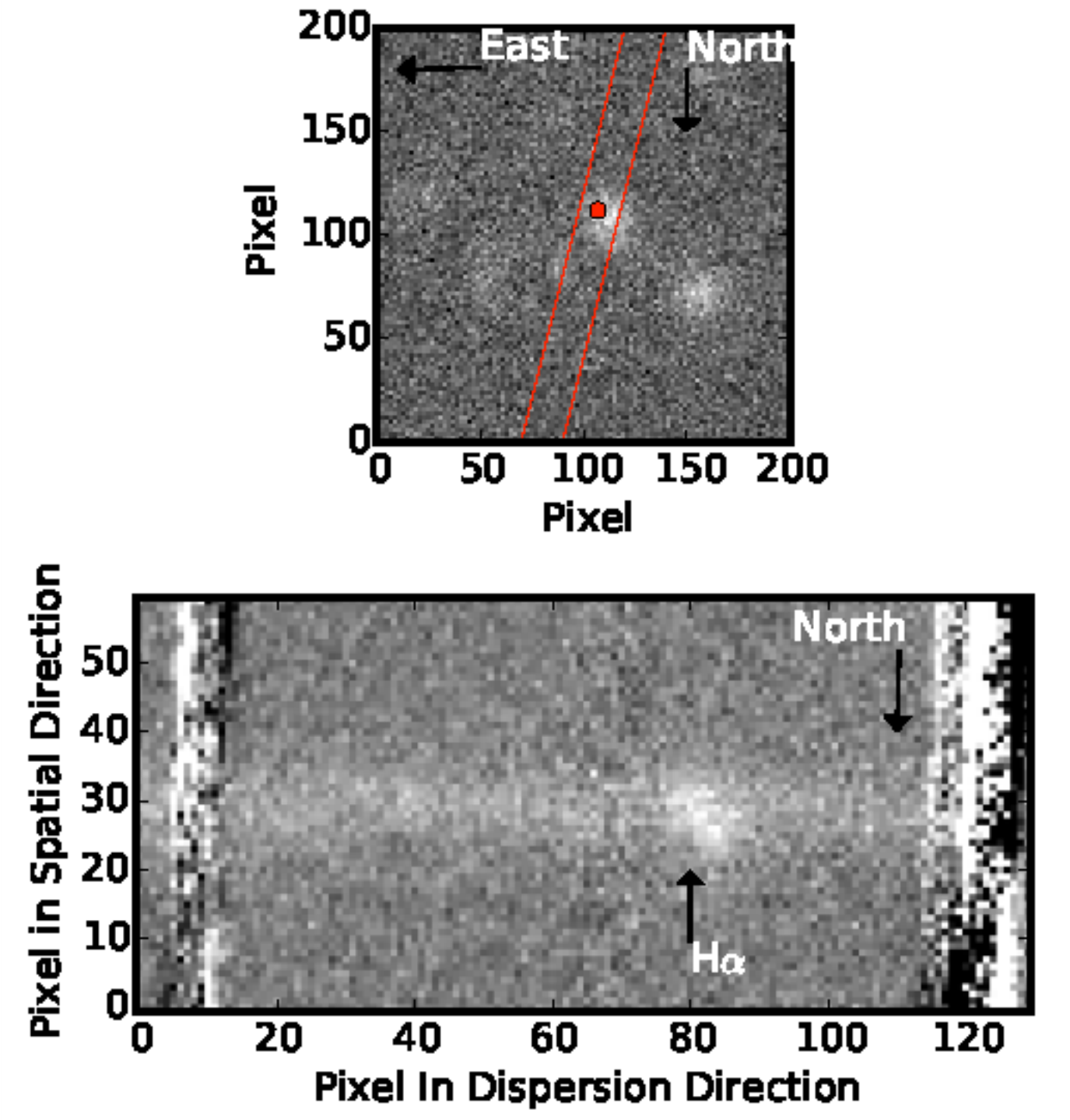}
\caption{The top panel shows the DEIMOS slit (red lines) with Position Angle (PA) of 200 overlaid on iPTF13ehe (red dot) and the host galaxy. The slit width is 1.2$^{''}$.
The bottom panel is the reduced 2-dimensional DEIMSOS spectrum taken on December 21, 2014. The x-axis is the dispersion direction, with a scale of 0.33\AA\ per pixel, and the y-axis is the spatial direction with a scale of 0.1185$^{''}$ per pixel. \label{deimos2d}}
\end{figure*}

The observed flux variations with time in the narrow H$\alpha$ line are likely due to the combination of the weather changes and different slit position angles.  The consistent redshift and SFR measured from [O\,II] and narrow H$\alpha$ suggest that the narrow H$\alpha$ line is mostly from the host galaxy and not from the supernova.  Additional support for this conclusion comes from the high resolution DEIMOS spectrum taken on December 21, 2014.  Figure~\ref{deimos2d} shows the reduced, 2-dimensional spectrum around the H$\alpha$ region. We found that the narrow H$\alpha$ component is resolved in velocity, and is consistent with the host galaxy rotation velocity field of 65\,km/s (roughly 1 spectral resolution, 1.8\AA). Furthermore, Figure~\ref{deimos2d} top panel shows the DEIMOS slit overlaid on the direct image of iPTF13ehe (red dot) and the host galaxy.  The narrow H$\alpha$  with extended velocity (the bottom panel) does not have any obvious strong emission corresponding to the spatial location of iPTF13ehe (south of the center of the host galaxy).  We therefore conclude that the narrow H$\alpha$ line is mostly likely from the host galaxy of iPTF13ehe.

The final piece of evidence supporting this conclusion is discussed in \S~4.2 below, where we argue that the H-rich CSM shell is likely to be mostly neutral because of the shell mass limit constraint by the Thomson optical depth $\leq1$ in the nebular phases.

\begin{figure*}[!htb]
%\plotone{/Users/lyan/Work/SLSN/13ehe/spec/nebcomp1v2.pdf}
\plotone{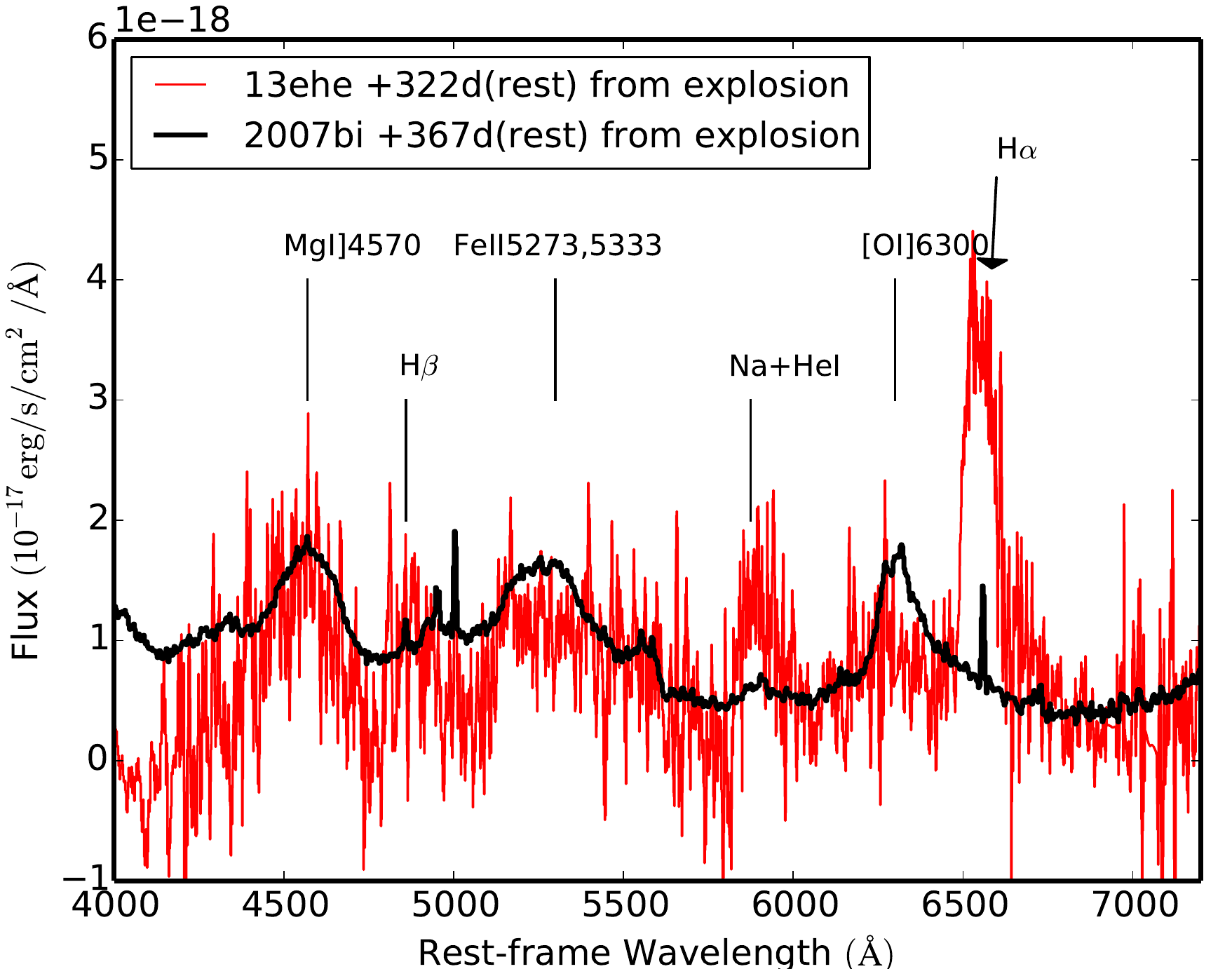}
\caption{ An optical nebular-phase spectrum of iPTF13ehe, taken $+322$days (rest-frame) from the estimated explosion date.  We compare this spectrum  with that of SN2007bi, which was taken at +367\,days (rest-frame) from the estimated explosion date, downloaded from WISeREP \citep{Gal-yam2009, Ofer2012}. \label{nebcomp}}
\end{figure*}

In the remaining of this section, we present our analysis of the broad H$\alpha$ component.
Figure~\ref{nebcomp} presents the first nebular spectrum ($t_{rest}^{exp}$\,=\,+322\,days) in comparison with the spectrum of SN2007bi taken at a similar phase. 
The strongest feature in the iPTF13ehe spectrum is its broad H$\alpha$ whereas that of SN2007bi has no traces of either H$\alpha$ or H$\beta$ from the supernova.  This broad  H$\alpha$ line is likely produced when the iPTF13ehe ejecta run into a H-rich CSM and the kinetic energy is converted into thermal emission, a part of which escapes in the H$\alpha$ and H$\beta$ lines. It is  intriguing that the emission from the [O\,I]\,6300,6363\,\AA\  is  very weak or absent, whereas this feature is very strong in SN2007bi.
The cooling ejecta from iPTF13ehe also produced emission such as broad Mg\,I]\,4570\,\AA,  the possible blend of Na\,5890\,\AA\ + He\,I\,5876\,\AA\ lines, and a blend of broad Fe\,II\,5169, 5261, 5273, 5333\,\AA.  The weak broad feature around 4861\,\AA\ could be H$\beta$, with a similar physical origin as that of H$\alpha$.

We note that the host galaxy star light was subtracted from the observed spectra as follows.  The host spectrum is constructed using a Bruzual-Charlot model \citep{Bruzual2003} constrained to have the same star formation rate (SFR) measured from the narrow [O\,II]\,3727\AA\ from the host galaxy.  The continuum decrement around 4000\AA\ is also used to match the model host spectrum.  Our data is inadequate to determine if a fraction of the narrow H$\alpha$ emission line is from the supernova. 

\begin{figure*}[!ht]
%\plotone{/Users/lyan/Work/SLSN/13ehe/spec/halphafit_fig9.pdf}
\plotone{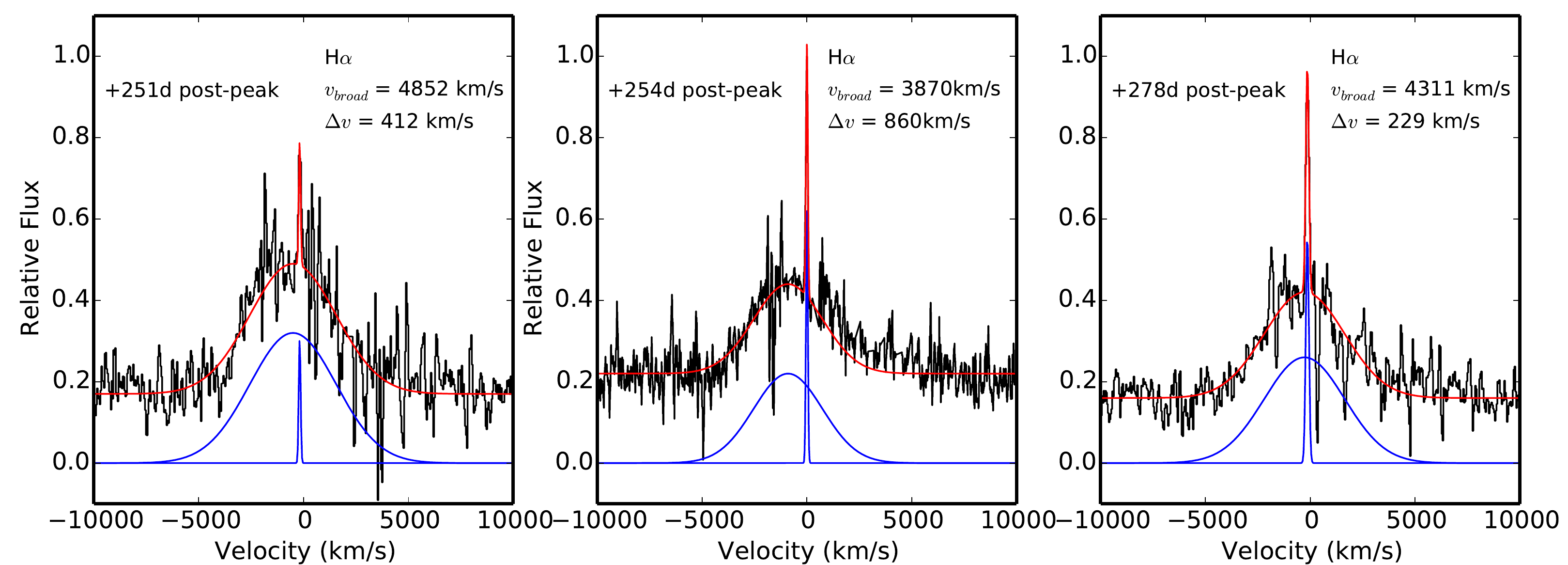}
\caption{The Gaussian profile fitting to the broad and narrow H$\alpha$ lines for the three epochs at +251, +254 and +278 days post-peak.  \label{spec-halpha}}
\end{figure*}

We fit gaussian profiles to both the broad and narrow H$\alpha$ lines in the spectra, shown in Figure~\ref{spec-halpha}.  The broad component has a FWHM between $3870 - 4850$\,km/s, which did not change much between December 17, 2014 and January 22, 2015. The integrated H$\alpha$ fluxes are $5.2-3.8$$\times10^{-16}$\,erg/s/cm$^2$, implying $L_{H\alpha}$\,=\,$2.2-1.6$$\times10^{41}$\,erg/s, decreasing by 20\%\ over a period of 60\,days. 
We also measured the velocity shifts between the narrow and broad components, $\delta v$\,$\sim$\,410 and 230\,km/s  for  the LRIS spectra taken on December 17, 2014 and January 22, 2015.  

Finally, we conclude that the broad component is from the SN ejecta interaction with a H-rich CSM, similar to the intermediate-velocity-width Balmer lines frequently observed among SN IIn. The $\sim$4000\,km/s of the broad component indicates the thermal, random motion of the shock ionized H atoms.  We assume that the velocity difference between the broad and narrow component, $\sim$300\,km/s, is the H-shell systematic velocity.  This assumption affects the calculations of the shell radius and when the shell is ejected by the progenitor star.  We speculate that the velocity shell can not be much larger than a few 100\,km/s, otherwise, the wavelength center of the broad component would be significantly shifted from the host galaxy redshift. However, it is possible that the shell could move slower than what we assumed.

%\begin{figure*}[!ht]
%\plotone{/Users/lyan/Work/SLSN/13ehe/spec/nebular5.pdf}
%\caption{The time evolution of the iPTF13ehe nebular spectra.  The SN emission features in the spectra taken a month apart are getting weaker. \label{nebevol}}
%\end{figure*}

\subsubsection{Nebular Emission Models}

A nebular emission model was computed for two epochs of the iPTF13ehe spectra (December 17, 2014 and Jan 22, 2015) using the code described in Mazzali et al. (2007). The code computes the heating of SN ejecta following the deposition of $\gamma$-rays and positrons from $^{56}$Ni and $^{56}$Co decay, balancing this by cooling via line emission in Non-Local Thermodynamic Equilibrium (NLTE).

\begin{figure}[!ht]
%\plotone{/Users/lyan/Work/SLSN/13ehe/spec/nebmodel.pdf}
\plotone{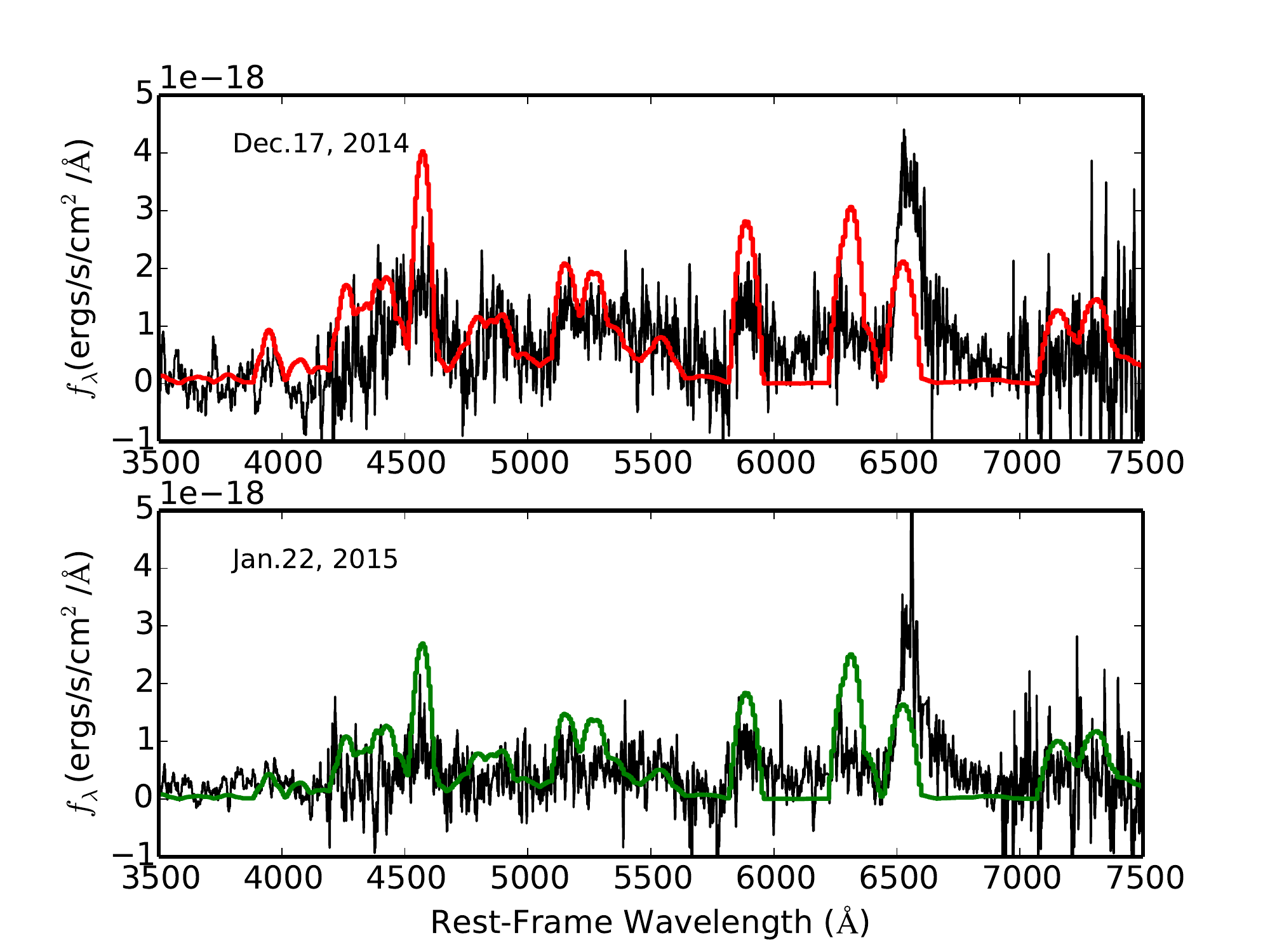}
\caption{The 1-D nebular emission line models are plotted against the spectra taken on two different epochs separated by a month. The colored spectra are model predictions, and the black lines are observed spectra. \label{nebmodel}}
\end{figure}

In the case of iPTF13ehe, given the low signal-to-noise ratio (SNR) of the nebular spectra,
line profiles could not be modeled, so a 1-D version of the code was used. The model spectra are compared with the observed data in Figure~\ref{nebmodel}.
Line width was matched to an ejecta velocity inside which all emission was assumed to
occur. This velocity is quite low, 4000\,km/s. The intensity of the Fe emission in
the blue was used to determine the mass of $^{56}$Ni, which depends also on
cooling from other species. The [O\,I]\,6300,6363\AA\ emission line is unusually
weak in iPTF13ehe in comparison with the nebular spectra of SN2007bi and other SN Ic events.
Ignoring all material at velocities above 4000\,km/s,  we obtain
a reasonable match to the spectra (excluding H$\alpha$) for M($^{56}$Ni)\,$\sim$\,$
2.5M_\odot$. This would correspond to a progenitor star of $\sim$\,95$M_\odot$ in the
models of Heger \&\ Woosley (2002). However, the oxygen mass in our model is only
13$M_\odot$, much less than the predicted value of 45$M_\odot$ \citep{Heger2002}. Although more
oxygen may be located at velocities above 4000\,km/s and not be significantly excited by radiation coming from the core. It is possible that a significant part of the CO core of the star was
lost before the explosion.  This is clearly in contradiction to the Heger \&\ Woosley (2002) model prediction.
In addition, we have kept the masses of Si and S at the values of the 95$M_\odot$
model of Heger \&\ Woosley (2002) (20 and 8$M_\odot$ respectively). This leads to
strong cooling lines of these elements, which are not in contradiction with the
optical data but are predicted to be very strong in the near-infrared, which is
unobserved. Should these elements be reduced in mass, the $^{56}$Ni mass would also
be reduced.   This naive comparison of model and data is to illustrate the obvious limitations and contradictions in the existing models of superluminous supernovae. In the case of iPTF13ehe, lack of any
[O\,I]\,6300\AA\ emission imposes a challenge to the models assuming massive progenitors.

%Despite these large uncertainties, it is reassuring that the same input produces
%spectra that match both epochs, as shown in Figure~\ref{nebmodel}. 
%%%% MOVE discussion on nebular model fitting to next section
%%%%
%We construct a nebular spectrum model to fit the observed nebular spectra, as shown in Figure~\ref{paolofit}.
%The $^{56}$Ni mass inferred from the model is 2.5$M_\odot$, the Zero-Age-Main-Sequence (ZAMS) mass is $\sim (95 - 100)M_\odot$. This is in the mass range where Pulsational Pair-Instability Supernova (PPISN) could work for 13ehe. 
%With the PPISN parameters from Heger \&\ Woosley (2002), our model predicts the ejecta mass of $\sim50 M_\odot$.  However, our model fails to produce as much as Oxygen as much as what H\& Woosley model predicts, $45M_\odot$. 

\section{Discussions }
\label{discuss}
 
\subsection{Physical Characteristics of the explosion}
The light curves and spectra provide measurements of $t_{rise}^{peak}$, $L_{bol}^{peak}$, and $v_{ej}$, which allow us to make the following physical parameter estimates.

We begin with the supernova ejecta mass $M_{ej}$.  The SN rise time is determined by how long it takes photons to radiatively diffuse out to the emitting surface.  This radiative diffusion time $t_{diff}$ can be derived based on electron Thomson scattering and expressed in terms of $M_{ej}$, $t_{diff}$\,=\,$f (\kappa M_{ej}/Rc)$, where $f$\,=\,${9 \over 4\pi^3}$ and $\kappa$ is mass opacity in cm$^2$g$^{-1}$ \citep{Arnett96, padmanabhan2000}. The characteristic time scale for a supernova LC is defined as $t_c$\,=\,$\sqrt{2t_ht_{diff}}$, where $t_h$ is hydrodynamic scale, $R/v_{ej}$. Thus, $t_c$\,=\,$\sqrt{2f\kappa M_{ej}/(c\times v_{ej})}$.  

%This radiative diffusion time scale $t_{diff}$ can be written as $(R/l_{mf})$$\times$$(R/c)$, with $l_{mf}=1/(\sigma_Tn)$ being the mean free path from electron Thompson scattering, $n$ the volume number density, $R$ the photospheric radius, and $c$ the speed of light. Expressed in terms of $M_{ej}$ instead of $n$, $t_{diff}$\,$\sim$\,$(\kappa M_{ej}/R^3)$$\times$$(R^2/c)$\,$ \sim$\,$\kappa M_{ej}/Rc$, where $\kappa$ is mass opacity in units of cm$^2$g$^{-1}$.  The precise equation is $t_{diff}$\,=\,$f (\kappa M_{ej}/Rc)$, with $f$\,=\,${9 \over 4\pi^3}$\citep{Arnett96, padmanabhan2000}. The characteristic time scale for a supernova LC is defined as $t_c$\,=\,$\sqrt{2t_ht_{diff}}$, where $t_h$ is hydrodynamic scale, $R/v_{ej}$. Thus, $t_c$\,=\,$\sqrt{2f\kappa M_{ej}/(c\times v_{ej})}$.  

This equation is the basis for an empirical scaling relation, $t_{rise}^{rest}$\,$\propto$\,$t_{c}$\,$\propto$\,$\sqrt{(M_{ej}/v_{ej})}$. Using this relationship and another well studied Type Ib event, SN\,2008D, with $v_{ej} = 10,000$\,km/s, $t_{rise}^{rest}$\,=\,19\,days, and $M_{ej}$\,=\,7$M_\odot$ \citep{paolo2008}, we derive $M_{ej}$(iPTF13ehe)\,=\,173$M_\odot$.  Using $t_{c}$\,$\simeq$\,$t_{rise}^{rest}$, we calculate $M_{ej}$\,$\sim$\,67\,-\,219$M_\odot$ assuming $\kappa$\,=\,0.1\,cm$^2$g$^{-1}$ for an ejecta composed of heavy elements.  The mass opacity ($\kappa$) due to electron scattering is determined by the ionization fractions of C, O and Fe, as well as line opacity. For ejecta composed of heavy elements, various studies have used $\kappa$ in the range of $(0.01 - 0.2)\,$cm$^2$g$^{-1}$ \citep{Arnett1982, bersten11, bersten12}.  The $M_{ej}$ estimate is quite sensitive to $\kappa$ and $t_c$, which are very uncertain with our current knowledge. Regardless the large uncertainty in $M_{ej}$, We can conclude that the exploding core mass of iPTF13ehe is $>$67$M_\odot$.  This implies that the progenitor star of iPTF13ehe must be very massive. 

%This is a factor of 4 less than that derived using the scaling relation.  Using $\kappa$\,=\,0.05\,cm$^2$g$^{-1}$ (ejecta with more heavy elements) would drive $M_{ej}$ larger by a factor of 2.  As discussed in \citet{Ofek2014}, the characteristic time scale for light curves may be better described by an exponential form, rather than defined here as the time interval between the peak date and the explosion date. As shown by Equation (6) in \citet{Ofek2014}, the exponential time scale $t_e=1.75 t_{rise}^{rest}$. Setting $t_e$\,=\,$t_{diff}$, we would have $M_{ej}$\,$\sim$102$M_\odot$.   The uncertainty in $M_{ej}$ based on our current understanding is quite large, a factor of 5. 
%The mass opacity ($\kappa$) due to electron scattering is determined by the ionization fractions of C, O and Fe, as well as line opacity. For ejecta composed of heavy elements, various studies have used $\kappa$ in the range of $(0.01 - 0.2)\,$cm$^2$g$^{-1}$ \citep{Arnett1982, bersten11, bersten12}.   We conclude that the exploding core mass of iPTF13ehe is at least $>$35$M_\odot$.  

The second physical parameter is the supernova kinetic energy, $E_{kin}$\,=\,${1 \over2}M_{ej} v_{ej}^2$\,=\,0.5$\times$(67\,-\,220)$M_\odot$$\times$$(13000)^2$\,km/s\,=\,(1\,-\,4)$\times 10^{53}$\,erg.  The implied kinetic energy $E_{kin}$ to the ejecta mass $M_{ej}$ ratio in the units of $10^{51}$\,erg per $1M_\odot$ is $\sim$1.6.  Comparing to the lower limit on the supernova radiative energy, $E_{rad}$$\geq$\,0.93$\times10^{51}$\,erg,  this implies $<$1\,\%\ of $E_{kin}$ being converted into visible radiation.  Most of the kinetic energy from this extreme power explosion is gone into expansion. 

Our inferred $E_{kin}$ is extreme in comparison with typical values observed among Type Ia and core-collapse SNe.  Specifically, for a normal core-collapse supernova, the total gravitational energy available from forming a neutron star is on an order of $10^{53}$\,erg, and a very large fraction of that is lost to free streaming neutrinos. So it is difficult to explain within a standard core-collapse model how iPTF13ehe could get such an extremely large kinetic energy. This imposes a challenge to models which use magnetars as energy sources.
One possible mechanism to produce such a large explosion energy is PPISN or PISN models for the most massive stars from Heger \&\ Woosley (2002).
However, we note that it is also possible that the ejecta mass $M_{ej}$ is overestimated by a factor of $(5-10)$.  As discussed earlier,  the commonly used methods have serious limitations, and it is not clear how to measure ejecta masses more accurately for SLSN events.

The third physical parameter is the $^{56}Ni$ mass.  If the power source for the observed luminous emission is radioactive decay of $^{56}$Ni, how much of this material would be required?  
The bolometric peak luminosity, 1.3$\times10^{44}$\,erg/s can constrain the amount of $^{56}$Ni and $^{56}$Co, based on $L_{bol}$\, =\,8$\times10^{42}$erg/s $\times M_{Ni}/M_\odot$ \citep{Arnett1982, smith2007b}.  The inferred $M_{Ni}$  is 16$M_\odot$. 
\citet{Katz2013} discussed another method, $\int_0^{t} Q(t') t' dt' = \int_0^{t} L_{bol}(t') t' dt'$, with $t$\,$>>$40\,days, $Q$ is the radioactive heating function ($^{56}$Ni$-$$> $$^{56}$Co$-$$>$Fe), and can be expressed as $Q(t) = 3.9\times10^{10} e^{-t/7.6e+5} + 6.78\times10^9 (e^{-t/9.59e+6} - e^{-t/7.6e+5})$. All are in cgs units. Using this relation and the $r$-band absolute magnitude LC, we derive a lower limit on $M_{Ni}$ of 13$M_\odot$. 
These numbers are much larger than those of any classical supernovae. The calculations assuming $^{56}Ni$ as the single power source yield much larger estimates than the prediction from our nebular emission model.  This may indicate that multiple power sources could be in play for producing the iPTF13ehe emission. Radioactive decay could be one of them, and the total $^{56}$Ni mass does not have to be so extremely large.
 
Finally, If the emission at the peak luminosity is considered as a blackbody,  
the energy can be estimated as $E_{rad}$\,=\,${4\pi \over3}R_{ph}^3$\,$\times$$aT_{eff}^4$, here $R_{ph}$ is the photospheric radius at the peak luminosity, and $T_{eff}$ is the blackbody effective temperature.  As discussed  in \S\ref{photo-spec}, $T_{eff}$ is  quite low, $7000$\,K.  With the lower limit on $E_{rad}$,  this equation implies the size of photosphere is $R_{ph}$\,$\sim$\,1.8$\times10^{16}$\,cm, a very large radius at peak luminosity compared to that of normal SNe.  This is mostly due to the low value of $T_{eff}$. It is not clear why some SLSNe-I seem to have much cooler temperatures at similar early phases than others.  

\subsection{What produced the broad H$\alpha$ emission in the late-time spectra?}
The key new result we present  here is the discovery of a broad H$\alpha$ line with a velocity width of 4000\,km/s in the late-time spectra.  The question is how this emission might be produced.  The simple explanation is  the interaction of the SN ejecta with a H-rich CSM, as commonly seen in SLSN-II spectra.  However, the key difference here is that this H-rich material is in a shell, located at a distance much further away from the explosion site.  The reason for a discrete shell rather than a continuous CSM material such as wind-driven mass loss is because the early-time spectra do not show any signatures of the ejecta-CSM interaction. This interaction only appears at late times.  

At the rest-frame $-9$\,days pre-peak, the SLSN ejecta velocity was 13,000\,km/s.  Let us take the estimated explosion date as 2456575.6\,days (the latest explosion date from the exponential fit).  The first detection of H$\alpha$ is on JD\,=\,2457008.5, giving an interval in the rest-frame $\Delta t_{rest}$=322\,days.   As Figure~\ref{abslc} shows that there is no spectroscopy between $+13$ and $+251$\,days (post-peak, rest-frame), therefore, the precise time when H$\alpha$ emission line first appears can not be determined.  The date of the third spectrum without H$\alpha$ (JD = 2456689.5) can be used to set the lower limit on $\Delta t_{rest}$\,=\,$85-322$\,days.  In practice, the lower limit value of $85$\,days is unlikely.
In the rest of the calculations, we use only $\Delta t_{rest}$\,=\,332\,days for the simplicity. 
Assuming that the ejecta did not slow down significantly before running into the CSM shell, the approximate distance travelled by the ejecta is $R_{rest}$\,=\,$v_{ej}\times \Delta t_{rest}$\,=\,4e+16\,cm.
The broad H$\alpha$ line has a width of $\sim$4000\,km/s, and is separated from the narrow H$\alpha$ by roughly ($400-230$)\,km/s. 
We interpret the 4000\,km/s line width as the thermal motion of shock-ionized hydrogen atoms, and the velocity shift of $\sim$300\,km/s  is $v_{csm}$, the CSM shell velocity.  We can set a limit on the time scale, $\Delta t$, when this material was ejected by the progenitor star using $\Delta t$$ \times$$ v_{csm}$\,=\,$R_{rest}$, giving $\Delta t^{rest} \leq 40$\,years. This is an upper limit since the CSM shell could initially have had higher speed.

Let us consider an H-rich CSM shell with a radius of 4$\times$$10^{16}$\,cm surrounding iPTF13ehe. This naturally begs two questions:
(1) is this material from the progenitor star or a part of the galactic ISM? (2) why does this CSM not produce any observable signatures --- either in emission or absorption --- in the early spectra when the UV photons from the early explosion would have interacted with this material?  

This H-rich CSM shell is only 4$\times$$10^{16}$\,cm from the location of the progenitor star, which is several orders of magnitudes smaller than a typical size of a H\,II region. Therefore, this H-rich CSM is probably not a part of the galactic ISM, but rather more likely produced by either wind mass loss or ejection due to some other mechanism, such as pulsational pair instability, occurring in massive stars ($>$67$M_\odot$).

As early as $-$\,days pre-peak, the follow-up spectra reveal strong supernova signatures, which implies that at this phase, the CSM shell must be optically thin 
to visible photons.  The fact that we also do not detect any narrow hydrogen recombination lines from the H-rich CSM shell, commonly seen in the spectra of SNIIn, could suggest two possibilities. First is that this H-shell has already become neutral when the first spectrum was taken.  Second is that this H-shell is ionized, but the early spectra are dominated by the supernova light, and the exposure times are too short to detect any H-recombination signals from this ionized shell.  However, as shown below, this second scenario is unlikely because of short recombination time scale.

When this CSM shell was initially ejected by the progenitor star 50\,years ago, the medium was very likely fully ionized. This ionized state was maintained by the heat sources from the progenitor star, and probably continued to the early phase of the supernova explosion.
However, when the supernova photosphere cools down,  there are no more heat sources  and the ionized H atoms in this shell will recombine.  This recombination time scale must be less than 62\,days (rest-frame, from the explosion date of 2456575.6) , {\it i.e.} the recombination is completed before the date of the first optical spectrum.  Assuming Case B recombination, the recombination time scale is $t_{rec} \sim 10^{13}/n$\,seconds $<62$\,days, with $n$ being the volume density, and $n$\,=\,$M_{csm}/4\pi m_HR^2_{rest} w$, $R_{rest}$\,=\,$4\times10^{16}$\,cm (radius of the shell), $m_H$ is the hydrogen mass and $w$ is the width of this shell.  Assuming the shell width $w$ is only 10\%\ of the shell radius, the above equations yield $M_{csm}>0.03M_\odot$.  

As shown below, the shell mass $M_{csm}$ is constrained to be $<$30\,$M_\odot$. At the upper limit of 30$M_\odot$, the H recombination time scale $t_{rec}$\,$\sim$\,$10^5$\,seconds. Thus, at the early phases of the spectral observations, the H-shell is already neutral.

Once this CSM shell becomes neutral, without any other heat sources, it would stay neutral until the SN ejecta run into this shell. The shock front generates high energy photons which ionize hydrogen atoms again. These ionized H atoms recombine, and produce the observed H$\alpha$ and H$\beta$ emission lines.  The important question is if this shock-ionized shell (or partially ionized shell) is optically thick to Thomson scattering. The fact that the late-time spectra do display SN spectral signatures, such as Mg\,I]\,4570\AA, suggests that this shock-ionized shell is probably not optically thick to Thomson scattering.  Other evidence that the CSM shell is not extremely dense comes from the {\it SWIFT} soft X-ray observation taken on December 23, 2014, yielding a $3\sigma$ limiting luminosity of 3$\times$$10^{43}$\,erg/s.  In addition, the ejecta interaction with extremely dense CSM would produce elevated continuum emission, which is not observed in the late-time absolute $g$-band light curve.

So if the shock ionized CSM shell is optically thin, we have $\tau_{thomp} = \kappa \rho w$\,$\leq1$, here $\rho$  is the density and $w$ the width of this CSM shell.  With $\rho = {M_{csm} \over 4\pi R^2 w}$, the above equation is simplified as $M_{csm} \leq {4\pi R^2 \over \kappa}$, independent of the width of this shell. Assuming $\kappa=0.34$\,cm$^{2}$/g for a H-rich medium, we have $M_{csm} \leq 30M_\odot$.  This mass value corresponds to a volume number density and a column density of 4$\times10^{8}$\,cm$^{-3}$ and $10^{24}$\,cm$^{-2}$ respectively. We predict that this shell should produce strong Ly$\alpha$ absorption features if any UV spectra were taken.  There should not be much $H\alpha$ absorption because without external excitation source, most of the H atoms are in the $n=1$ ground state.  To have significant population in the $n=2$ state with collisional excitation, it would require the gas to have very high temperature, such that $KT \sim \Delta E_{21} = 10$\, ev$\sim$100,000\,K.  The shell around iPTF13ehe is unlikely to have such a high temperature.

With the estimated $M_{csm}$, we can calculate the kinetic energy of this CSM shell, ${1\over 2}M_{csm}v_{csm}^2=2\times10^{49}$\,erg/s. Here we assume that the CSM shell systematic velocity is $\sim$300\,km/s, roughly the velocity shift observed between the broad and narrow H$\alpha$ lines.  The initial $v_{csm}$ could be larger.  The energetics of this shell is within the predictions of PPISN models \citep{Woosley2007}.

We note the CSM shell may be partially ionized. This would lead to higher $M_{csm}$ value and suggests that collisional excitation might be important for producing H$\alpha$. The shock heated CSM shell with Balmer dominated emission is a complex system.  The broad H$\alpha$ emission is likely produced by charge exchange between fast moving neutral H-atoms \citep{Chevalier1980,Morlino2013}.  Better quantitative estimates will need to apply the theory of Balmer dominated emission shock. This is beyond the scope of this paper. 
%However, if all of the H$\alpha$ luminosity was from hydrogen recombination,  we would have $L_{H\alpha} \leq 5.6\times10^{43} {M_{csm}^2(M_\odot) \over r_{15}^2 w_{15}} $\,ergs/s, here $r_{15}$ and $w_{15}$ is the shell radius and the width in units of $10^{15}$\,cm \citep{Ofek2013}.  With the observed $L_{H\alpha}=2\times10^{41}$\,ergs/s,  the limit is $M_{csm} > 2.3M_\odot$.  This is consistent with the upper limit $M_{csm}$$\leq29.6M_\odot$.

%We note that the assumptions used in the above calculations might not hold. This CSM shell might be partially ionized. This would lead to higher $M_{csm}$ value and suggests that collisional excitation might be important for producing H$\alpha$. These complicated issues need to be addressed in future more detailed studies.

\subsection{iPTF13ehe: pulsational pair instability supernova versus other models}
Detection of a H-rich CSM shell around a H-poor SLSN is a natural prediction from the PPISN model \citep[e.g.][]{Woosley2007,Waldman2008}.   Table 1 in the supplemental material from \citet{Woosley2007}  illustrates that in the cases of He-core masses greater than $53M_\odot$, the time interval between the first and the second instability pulses could be as long as $6\times10^9$\,seconds, and the subsequent instabilities would happen more frequently.  iPTF13ehe lost its H-shell about 40\,years ($10^9$\,seconds, rest) before the SN explosion.  In addition,  the kinetic energy of this CSM shell is about $2\times10^{49}$\,erg/s, similar to what is predicted in Table 1 of  \citet{Woosley2007}.   Both the time scale and energetics support the hypothesis that iPTF13ehe could be a PPISN candidate.
In this scenario, iPTF13ehe started with a progenitor star with an initial mass of $>70M_\odot$, ejected about $<$30$M_\odot$ H-envelope about 40\,years ago (rest) during the first episode of pulsational pair instability.  Following the first mass ejection, there is at least one, possibly more pulsational pair instabilities before the supernova explosion.  The later ejected H-poor CSM shell tends to be faster and more energetic than the previous one, naturally leading to shell-shell collision \citep{Woosley2007}.  The reason for possible additional instability pulses is  that H-poor CSM shell-shell collision could provide one of the power sources for the observed LC and spectral features. 

However, this PPISN model could have one potential problem.  The highest kinetic energy generated by these pulsational pair instabilities is predicted to be  8$\times10^{50}$\,erg, no more than $10^{51}$\,erg. The relative kinetic energy between the 2nd and 3rd pulses would be even smaller.  Therefore, even with the most efficient kinetic and thermal energy conversion, it may be hard to produce the peak radiative energy measured for iPTF13ehe ($>$9.3$\times10^{50}$\,erg).  One solution proposed in \citet{quimby2011} is that after several episodes of pulsational pair instabilities, the core undergoes supernova explosion. 
The ejecta interaction with the last H-poor CSM shell could provide a more energetic reservoir for powering the observed emission.  This is derived from an earlier idea proposed for SLSNe-II by \citet{Woosley2007}.

The second caveat regarding the PPISN model is that the spectra of iPTF13ehe do not show much [O\,I]\,6300\AA\ emission in all of the phases we had data. The iPTF13ehe ejecta seems not to have much oxygen material.  This is clearly in contradiction with the models calculated by \citep{Heger2002}, which predicts an oxygen dominated core with the mass $>$50$M_\odot$. Our estimated ejecta mass has a lower limit of 70$M_\odot$, suggesting a very massive core. However, this calculation depends on the assumed value of opacity, $\kappa$. If adopting a higher value of 0.2, the ejecta mass would be a factor 2 smaller. 
These uncertainties may suggest a less massive progenitor star which does not go through pulsational pair instabilities. For example, a Luminous Blue Variable (LBV) could eject a massive H envelope from the progenitor star during its instability episode, like Eta Carina. Then a massive core collapse model, such as the one proposed by Moriya et al. (2010), is needed to explain the energy output of iPTF13ehe. 

%The key to the PPISN model ability to explain the high energy output and the H-poor spectra observed in SLSNe-I is interaction, whether it is shell-shell collision or ejecta-shell interaction. It is very efficient in transforming abundant kinetic energy into radiative energy. None-interaction model, such as massive core collapse model, was also proposed by \citet{Moriya2010}, where after the initial hydrogen envelope is ejected, the massive C/O core collapses without any H-poor CSM shells.  One particular restriction of this model is that the ejecta mass has to be less than 50$M_\odot$, otherwise, the C/O core would be heavier than $50M_\odot$, inevitably leading to pair instability.  The observations of iPTF13ehe and SN2007bi set fairly stringent lower limits on the ejecta masses, higher than $50M_\odot$.  This implies that this none-interaction model is likely ruled out. 
 
With very low metallicity and rotation, 
a variation of the PPISN model could have a progenitor star with a much lower mass than $95-150$\,$M_\odot$ predicted in the Woosley et al. (2007) study. As pointed out by studies of \citet{Chatzopoulos2012,Yoon2012}, low metallicity and rotating stars could undergo pulsational pair instabilities at initial stellar masses as low as $\sim$\,$50-70$$M_\odot$.  Our constraint on the iPTF13ehe progenitor mass is not very strigent, the lower limit is less than the initial mass of $95M_\odot$ predicted by the Woosley 2007 model. If low metallicity and rotation are relevant, iPTF13ehe could be a PPISN candidate.

Finally, there is another alternative physical model, which was briefly mentioned in \citet{Woosley2007}.  A $95-150$$M_\odot$ star with rotation and magnetic torques would initially evolve in a similar fashion to one without rotation and magnetic field. It will undergo episodes of pulsational pair instability which eventually produce a C/O core with a mass in a range of $40-60$$M_\odot$.  However, the difference is that the rotating star with magnetic torque can end up forming a neutron star with a fast spin-period of a few milli-seconds and magnetic field strength of $10^{15}$\,gauss -- {\it i.e.} a magnetar \citep{Duncan1992,Heger2005}. The spin-down of a magnetar can  provide sufficient power for a SLSN, as shown in several studies \citep{Kasen2010,Nicholl2013,Inserra2013}. For iPTF13ehe, it is possible that its massive progenitor star experiences several episodes of pulsational pair instabilities, ejecting several shells, and the final supernova explosion leaves behind a magnetar. The power sources for the observed LC and early-time H-poor spectra could well be a combination of a magnetar and the collision between H-poor CSM shells.  The magnetar scenario clearly needs more scrutiny in the future.

\subsection{How common are the SLSNe-I with late-time Balmer emission lines ?}
\begin{figure*}[!ht]
%\plotone{/Users/lyan/Work/SLSN/13ehe/10aagc/10aagcspec_v5.pdf}
\plotone{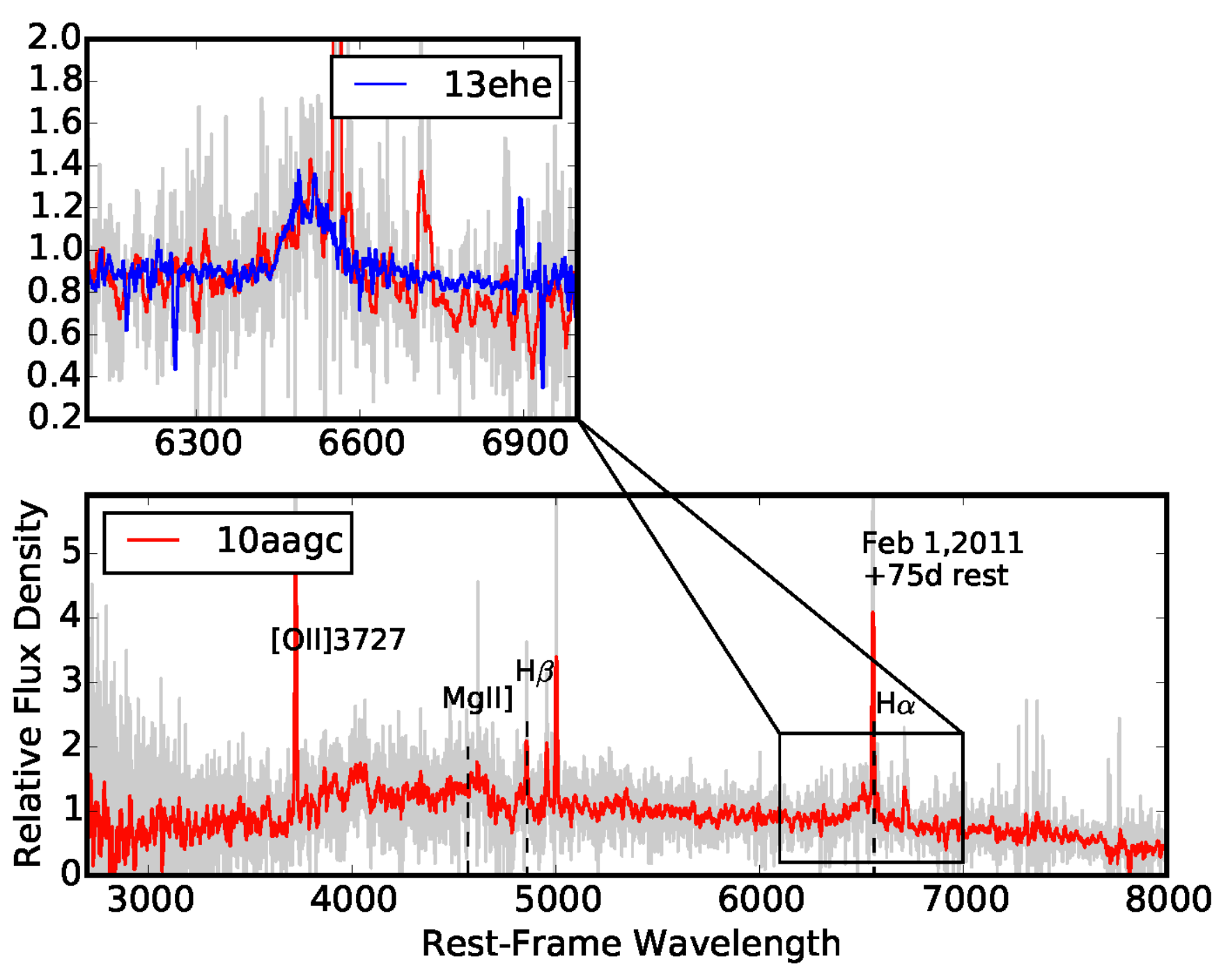}
\caption{The late-time spectrum of PTF10aagc. We overlaid the smoothed spectrum (red line) on the top of the original data (grey line). The smoothing length is 13\,pixels, corresponding to 15\AA.  In the zoom-in panel, we compare the broad H$\alpha$ component from iPTF13ehe (blue line) to that from PTF10aagc (red line). \label{10aagcspec}}
\end{figure*}

If PPISN is a possible model to explain iPTF13ehe, a related question is how common such an event is among all SLSNe-I. This question is difficult to answer because it depends on when spectroscopic observations are taken.  Of the 23 SLSNe-I at $z$\,$<$\,0.4 PTF discovered during 2009-2013, 13 events have at least one spectrum taken after 100\,days post-peak.  Of these 13 events, we found two cases with Balmer emission lines in the late-time spectra. The second case is PTF10aagc, a H-poor SLSN at $z$\,=\,0.207.  Figure~\ref{10aagcspec} shows the spectra taken at the phase of +75\,days post-peak (rest-frame), revealing a broad H$\alpha$ and the corresponding weak, but detected H$\beta$.  PTF10aagc may be another case of a SLSN-I with ejecta interaction with a H-rich CSM at late times, although in many ways PTF10aagc is different from iPTF13ehe.  Detailed discussion of this object is included in Quimby et al. (in preparation).  Our data suggests that at least 15\%\ of all SLSNe-I have late-time ($>$100\,days post peak) Balmer emission lines from ejecta interaction with H-rich CSM.   It is possible that much higher fraction of SLSNe-I would eventually show late-time spectral signatures of interaction with H-rich CSM.
However, the answer must depend on the mass loss mechanism the SLSN-I progenitor stars have.

%We conclude that of all H-poor SLSNe, probably about 15\%\ have late-time ($>$100\,days post peak) Balmer emission lines from ejecta interaction with H-rich shells. 

%Of the 27 hydrogen-poor SLSNe discovered by PTF and iPTF throughout 2009\,-\,2014, we found two cases with late time spectral signatures of ejecta interaction with a H-rich CSM. The second case is PTF10aagc, which is at $z$\,=\,0.207.  Figure~\ref{10aagcspec} shows three spectra taken at the phases of +3.6, +75 and 119\,days post-peak (the LC peaks at JD\,=\,2455503.6\,days).  PTF10aagc is classified as an hydrogen-poor SLSN based on its early phase spectrum with a steeply rising blue continuum and OII absorption features near 4500\AA, similar to those of SLSN-I PTF09cnd.  Th early spectrum (+3.6\,days post-peak, Figure~\ref{10aagcspec}) also detected a broad feature near 6563\AA. We compared this spectrum with that of a SN II at the similar phase, we conclude that the broad feature around 6563\AA\ is not H$\alpha$ because it does not have the expected broad H$\beta$. This un-identified feature might be a mixture of Si\,II and C\,II\,6578,6583\AA\ \citep{quimby2011,Hatano2001}.  At the late time, +75\,days post-peak, the spectrum shows a broad H$\alpha$ and the corresponding weak, but detected H$\beta$.  PTF10aagc may be another case of a SLSN-I with ejecta interaction with a H-rich CSM at late times, although in many ways PTF10aagc is different from iPTF13ehe.  For PTF10aagc, this H-shell was expelled by the progenitor star about 10\,years ago before the supernova explosion.
 
\section{Summary}
iPTF13ehe shows photometric and spectroscopic properties of a SLSN-R, similar to SN2007bi. The key characteristics are the long rise time ($83-148$\,days) and slow decay rate (0.0149\,magnitude/day), different from many SLSNe-I \citep{quimby2011}. The slow, linear decline of the rest-frame $g$-band LC does not completely rule out the radioactive decay as a possible power source because we do not have the proper late-time bolometric LC, which is required to make a meaningful comparison with the $^{56}$Co decay rate of 0.00977\,mag/day. Another feature which distinguishes iPTF13ehe from most SLSNe-I is its low blackbody temperature at the peak brightness, implying a very large photospheric radius.
We measured the peak bolometric luminosity of $1.3\times10^{44}$\,erg/s and ejecta velocity of 13000\,km/s.  The inferred ejecta mass is very large, in the range of $70-220M_\odot$, implying a very massive progenitor star regardless of the details of explosion physics.  The energetics of iPTF13ehe is in the extreme with the radiative energy $\sim$$10^{51}$\,erg and the supernova kinetic energy of $>10^{53}$\,erg, posing strong challenges to standard core-collapse models.  The derived kinetic energy $E_{kin}$ to the ejecta mass $M_{ej}$ ratio in the units of $10^{51}$\,erg/1$M_\odot$ is $\sim$1.6.

The new discovery from the iPTF13ehe observations is the detection of a broad H$\alpha$ emission line with a velocity of 4000\,km/s in its nebular phase spectra taken at $251-278$\,days. The late-time appearance of H$\alpha$ emission in iPTF13ehe is unique, very different from SN2008es \citep{Miller2009,Gezari2009} and CSS121015:004244+132827 \citep{Benetti2014}.  For these two superluminous events, although their early-time spectra revealed no traces of hydrogen, the broad H$\alpha$ emission lines were detected after $\sim$$+40$\,days post-peak.  It is likely that hydrogen exists in the photosphere, but is probably mostly ionized due to very hot temperature. Only after $+40$\,days post-peak when the photosphere cools down, the H-recombination lines start to appear. 

The situation in iPTF13ehe is quite different.  There are no hydrogen features at all in all of the early-time spectra, even when the temperature measured from the spectra is quite cool (7000\,K at the peak).  A broad and strong H$\alpha$ emission line only emerged in the late-time nebular spectra.  Independent of explosion models, this observation reveals the existence of a discrete and distant H-rich shell, which must have been expelled from the progenitor star some years ago before the supernova explosion.  The estimated shell mass and the associated kinetic energy of $10^{49}$\,erg/s suggest that the violent mass loss episodes are extremely energetic, able to unbind the entire hydrogen envelope.  One model which predicts such energetic mass losses is Pulsational Pair Instability Supernova (PPISN) for a star with the initial mass of $95-150$$M_\odot$ \citep{Woosley2007}. 

The results from iPTF13ehe suggest that future surveys of SLSNe at low redshifts ($z<0.4$) need to have well designed plans for the late-time follow-up observations, particularly at the nebular  phase. Any desire to measure statistics of PPISN candidates like iPTF13ehe would require more systematic follow-up observations than what have been done so far.

\acknowledgments
We thank the anonymous referee for the positive and constructive suggestions, which have helped to improve the paper. 
We benefited from discussions with Nick Scoville and Orly Gnat on collisional excitations in ISM.
We thank Mansi Kasliwal, Thomas Prince and Howard Bond for helping us to obtain the P200 photometry at one epoch.  Vicki Toy and John Capone from University of Maryland are acknowledged for taking the photometry observation using LMI on DCT.
A.G.Y. is supported by the EU/FP7 via ERC grant no. 307260, the Quantum Universe I-Core program by the Israeli Committee for planning and budgeting and the ISF; by Minerva and ISF grants; by the Weizmann-UK ``making connections" program; and by Kimmel and ARCHES awards. The Dark Cosmology Centre is funded by the Danish National Research Foundation.
This paper made use of Lowell Observatory's Discovery Channel Telescope (DCT).
Lowell operates the DCT in partnership with Boston University, Northern Arizona University, the University of Maryland, and the University of Toledo. Partial support of the DCT was provided by Discovery
Communications. Large Monolithic Imager (LMI) on DCT was built by Lowell Observatory using funds from the National Science Foundation (AST-1005313).
LANL participation in iPTF is supported by the US Department of Energy as a part of the Laboratory Directed Research and Development program. A portion of this work was carried out at the Jet Propulsion Laboratory
under a Research and Technology Development Grant, under contract with the National Aeronautics and Space Administration. Copyright 2015 California Institute of Technology.  All Rights Reserved. US Government support is acknowledged.
This research has made use of the NASA/IPAC Extragalactic Database (NED) which is operated by the Jet Propulsion Laboratory, California Institute of Technology, under contract with the National Aeronautics and Space Administration.  
Some of the data presented herein were obtained at the W.M. Keck Observatory, which is operated as a scientific partnership among the California Institute of Technology, the University of California and the National Aeronautics and Space Administration. The Observatory was made possible by the generous financial support of the W.M. Keck Foundation. The authors wish to recognize and acknowledge the very significant cultural role and reverence that the summit of Mauna Kea has always had within the indigenous Hawaiian community.  We are most fortunate to have the opportunity to conduct observations from this mountain.

{\it Facilities:} \facility{Palomar}, \facility{Keck}, \facility{SWIFT}, \facility{Discovery Channel Telescope}.

%\LongTables
\begin{deluxetable*}{ccccccccc}
\tabletypesize{\footnotesize}
\tablewidth{0pt}
\tablecaption{The $g$, $r$ and $i$-band Photometry
\label{tab:phot}}
\tablehead{
\colhead{Julian Date} &
\colhead{$r\tablenotemark{a}$} &
\colhead{$\sigma$}  &
\colhead{Julian Date} &
\colhead{$g$} &
\colhead{$\sigma$}  &
\colhead{Julian Date} &
\colhead{$i$} &
\colhead{$\sigma$} \\
% units of columns
%\hline \\
days & mag & mag & days & mag & mag & days & mag & mag \\
}
\startdata
2456495.2  &   22.68   &    99.9\tablenotemark{b} & 	2456662.7   &  20.08    &   0.06	&   2456662.7   &  20.05    &   0.17\\
2456577.7  &   22.43   &    0.48     &	2456667.9   &  20.06    &   0.08	&   2456663.6   &  19.75    &   0.05\\
2456588.0  &   21.01   &    0.35     & 	2456673.6   &  20.21    &   0.21	&   2456671.7   &  19.70    &   0.09\\
2456621.8  &   20.39   &    0.19     &	2456680.6   &  20.12    &   0.04	&   2456672.7   &  19.83    &   0.14\\
2456627.0  &   20.25   &    0.12     &	2456685.6   &  20.15    &   0.04	&   2456673.6   &  19.76    &   0.13\\
2456639.7  &   19.97   &    0.14     &	2456697.6   &  20.34    &   0.07	&   2456680.6   &  19.77    &   0.07\\
2456639.7  &   20.09   &    0.27     &	2456703.0   &  20.37    &   0.03	&   2456685.6   &  19.65    &   0.06\\
2456640.8  &   19.96   &    0.10     &	2456703.7   &  20.50    &   0.07	&   2456697.6   &  19.71    &   0.04\\
2456647.9  &   19.75   &    0.14     &	2456710.7   &  20.50    &   0.04	&   2456703.0   &  19.71    &   0.03\\
2456648.8  &   19.88   &    0.06     &	2456732.7   &  20.91    &   0.17	&   2456703.7   &  19.79    &   0.04\\
2456662.7  &   19.69   &    0.06     &	2456733.6   &  20.87    &   0.12	&   2456710.7   &  19.79    &   0.04\\
2456667.9  &   19.87   &    0.10     &	2456735.7   &  20.79    &   0.08	&   2456721.7   &  19.73    &   0.06\\
2456675.8  &   19.60   &    0.19     &	2456736.7   &  20.93    &   0.05	&   2456729.1   &  19.83    &   0.12\\
2456680.6  &   19.74   &    0.05     &	2456737.8   &  20.97    &   0.10	&   2456733.4   &  19.81    &   0.09\\
2456685.6  &   19.79   &    0.05     &	2456745.8   &  21.24    &   0.11	&   2456737.2   &  19.88    &   0.08\\
2456697.6  &   19.81   &    0.05     &	2456952.1   &  23.87    &   0.08	&   2456745.3   &  19.95    &   0.11\\
2456703.0  &   19.83   &    0.03     &	2456981.1   &  24.15    &   0.08	&               &           &       \\
2456703.7  &   19.77   &    0.04     &	2457044.9   &  24.65    &   0.08	&               &           &        \\
2456710.7  &   19.83   &    0.03     &	2457104.5   &  24.77    &   0.10	&               &           &        \\
2456721.8  &   19.98   &    0.07     &	            &           &   	        &                &          &        \\
2456727.6  &   20.06   &    0.09     &	            &           & 	        &               &           &        \\
2456728.6  &   20.28   &    0.12     &               & 	        &	        &              &           &          \\
2456729.6  &   20.08   &    0.10     &              &           &		&               &          &          \\
2456730.6  &   20.13   &    0.10     &              &           &		&               &           &         \\
2456731.6  &   20.01   &    0.06     &              &           & 		&                &          &          \\
2456732.6  &   20.11   &    0.08     &              &           &		&                &           &         \\
2456733.6  &   20.10   &    0.05     &	            &           &	        &                &           &          \\
2456735.7  &   20.20   &    0.07     &	            &           &	        &                &          &           \\
2456736.7  &   20.19   &    0.05     &	            &           &	        &                &          &            \\
2456737.8  &   20.13   &    0.07     &	            &           &	        &                 &          &           \\
2456745.8  &   20.56   &    0.09     &	            &           &	        &                 &          &          \\
2456911.0  &   21.97   &    0.05     &	            &           &	        &                 &           &          \\
2456952.1  &   22.36   &    0.07     &	            &           &	        &               &            &           \\
2456981.1  &   22.66   &    0.07     &	            &           &	        &                &          &            \\
2457044.9  &   23.18   &    0.07     & 	            &           &	        &              &           &             \\
2457071.7  &   23.35   &    0.08     &	            &           &	        &               &            &          \\
2457104.5  &   23.63   &    0.10     &	            &           &	         &             &             &          \\
\enddata
\tablenotetext{a}{The magnitudes include light from both the host and iPTF13ehe, and are in AB system.  
The host galaxy and the supernova are well separated in the HST images. The host galaxy $r$-band brightness is $24.24$, measured from the HST photometry. The host $g$-band magnitude is set to 24.9 in the paper. The analysis in this paper does not use any late-time $i$-band photometry, thus the host subtraction is not critical. $i$-band host galaxy photometry has not been measured. All errors are in $1\sigma$.}
\tablenotetext{b}{This $r$-magnitude is a 3$\sigma$ limit.}
\end{deluxetable*}

\clearpage

\begin{deluxetable*}{ccccc}
\tabletypesize{\footnotesize}
\tablewidth{0pt}
\tablecaption{The Spectroscopic Observation Log
\label{tab:obslog}}
\tablehead{
\colhead{Obs.Date} &
\colhead{Julian Date} &
\colhead{Instrument} & 
\colhead{Exp.Time$^a$} & 
\colhead{Inst. Res.$^b$} \\
% units of columns
%\hline \\
 & days &  & seconds & \AA \\      
}
\startdata

01/01/2014 & 2456658.5 & Keck/DEIMOS & 600 & 4 \\
01/06/2014 & 2456663.8 & P200/DBSP  &  1800 & 4.4 \\
02/01/2014 & 2456689.9 & P200/DBSP  & 1200  & 6.1  \\
12/17/2014 & 2457008.4 & Keck/LRIS    & 3100(blue), 2700(red) & 5.6\\
12/21/2014 & 2457012.4 & Keck/DEIMOS & 6000 & 1.8 \\
01/22/2015 & 2457044.4 & Keck/LRIS  &    3800(blue), 1800(red) & 5.6  \\
\enddata
\tablenotetext{a}{Keck/LRIS blue and red side exposure times are different.}
\tablenotetext{b}{Instrument spectral resolution is Full Width at Half Maximum (FWHM) measured from unresolved sky lines.}
\end{deluxetable*}

\end{document}